\documentclass[preprint]{emulateapj}
\usepackage{aas_macros}

\usepackage{graphicx}
\usepackage{amsmath,amssymb}
\usepackage{dcolumn}
\usepackage{bm}
\usepackage{comment}
\usepackage{natbib}
\newcommand{\beq}{\begin{equation}}
\newcommand{\beqa}{\begin{eqnarray}}
\newcommand{\eeq}{\end{equation}}
\newcommand{\eeqa}{\end{eqnarray}}

\newcommand{\simgt}{\lower.5ex\hbox{$\; \buildrel > \over \sim \;$}}
\newcommand{\simlt}{\lower.5ex\hbox{$\; \buildrel < \over \sim \;$}}

\newcommand{\bd}[1]{\mbox{\boldmath $#1$}}

\slugcomment{Submitted to ApJ}
\shortauthors{Osato, Shirasaki \& Yoshida}
\shorttitle{Impact of baryonic processes on weak lensing cosmology}

\begin{document}

\title{Impact of Baryonic Processes on Weak Lensing Cosmology:\\
Power Spectrum, Non-Local Statistics, and Parameter Bias}

\author{
Ken Osato$^{1}$\altaffilmark{\dag},
Masato Shirasaki$^{1}$ and
Naoki Yoshida$^{1,2,3}$
} 
\affil{
$^{1}$Department of Physics, School of Science, The University of Tokyo, 
7-3-1 Hongo, Bunkyo, Tokyo 113-0033, Japan\\
$^{2}$Kavli Institute for the Physics and Mathematics of the Universe (WPI), 
Todai Institutes for Advanced Study, The University of Tokyo,
Kashiwa, Chiba 277-8583, Japan\\
$^{3}$CREST, Japan Science and Technology Agency, 4-1-8 Honcho, Kawaguchi, 
Saitama, 332-0012, Japan
}
\altaffiltext{\dag}{E-mail: ken.osato@utap.phys.s.u-tokyo.ac.jp}

\begin{abstract}
We study the impact of baryonic physics on 
cosmological parameter estimation with weak lensing surveys.
We run a set of cosmological hydrodynamics simulations 
with different galaxy formation models.
We then perform ray-tracing simulations through the total matter density field
to generate 100 independent convergence maps of 25 ${\rm deg}^2$ field-of-view,
and use them to examine the ability of the following 
three lensing statistics as cosmological probes; 
power spectrum, peak counts, and Minkowski functionals.
For the upcoming wide-field observations such as Subaru Hyper Suprime-Cam (HSC) survey 
with a sky coverage of 1400 ${\rm deg}^2$,
these three statistics provide tight constraints on the matter density,
density fluctuation amplitude, and dark energy equation of state, but
parameter bias is induced by the baryonic processes such as gas cooling
and stellar feedback.
When we use power spectrum, peak counts, and Minkowski functionals,
the magnitude of relative bias in the dark energy equation 
of state parameter $w$ is at a level of, respectively, 
$\delta w \sim0.017$, $0.061$, and $0.0011$.
For HSC survey, these values are smaller than the statistical errors
estimated from Fisher analysis. The bias can be significant when the statistical 
errors become small in future observations with a much larger survey area.
We find the bias is induced in different directions 
in the parameter space depending on the statistics employed.
While the two-point statistic, i.e. power spectrum, yields robust results
against baryonic effects, the overall constraining power is weak 
compared with peak counts and Minkowski functionals.
On the other hand, using one of peak counts or Minkowski functionals,
or combined analysis with multiple statistics, 
results in biased parameter estimate.
The bias can be as large as $1\sigma$ for HSC survey, 
and will be more significant for upcoming wider area surveys.
We suggest to use an optimized combination so that
the baryonic effects on parameter estimation are mitigated. 
Such `calibrated' combination can place stringent {\it and} robust 
constraints on cosmological parameters.
\end{abstract}



\section{INTRODUCTION}

An array of recent observations of the large-scale structure of the universe 
such as cosmic microwave background (CMB) anisotropies
\citep[e.g.,][]{Hinshaw2013, 2014A&A...571A..16P}
and galaxy clustering
\citep[e.g.,][]{2010MNRAS.404...60R, 2014MNRAS.443.1065B}
established the standard cosmological model called $\Lambda$CDM model.
In $\Lambda$CDM model,
the energy content of the present-day universe is
dominated by two mysterious components: dark energy and dark matter.
Dark energy realizes the cosmic acceleration at present and
dark matter plays an important role of formation of rich structure in the universe.
However, we have not understood yet the nature of dark energy
and the physical properties of dark matter.
In order to reveal the mysterious dark components in the universe,
several observational programs are proposed and still under investigation.
Such observational programs include
Subaru Hyper Suprime-Cam (HSC)\footnotemark[1], 
the Dark Energy Survey (DES)\footnotemark[2],
and the Large Synoptic Survey Telescope (LSST)\footnotemark[3].
\footnotetext[1]{\rm{http://www.naoj.org/Projects/HSC/index.html}} 
\footnotetext[2]{\rm{http://www.darkenergysurvey.org/}}
\footnotetext[3]{\rm{http://www.lsst.org/lsst/}}
Space missions such as Euclid\footnotemark[4]
and WFIRST\footnotemark[5] are also promising.
\footnotetext[4]{\rm{http://www.euclid-ec.org/}}
\footnotetext[5]{\rm{http://wfirst.gsfc.nasa.gov/}}
Gravitational lensing is expected to be the main subject
of these future surveys that are aimed at studying
the large-scale structure of the universe at present
and in the past.

Weak gravitational lensing (WL) by large-scale structure in the universe
is the promising probe into properties of dark matter and dark energy \citep[for a review, see][]{Bartelmann2001,Munshi2008,Kilbinger2014}.
WL causes small distortion of image of distant source galaxies,
called cosmic shear, which reflect directly
the intervening matter distribution along a line of sight.
Recent cosmic shear observations have proved WL measurement to be a powerful tool
for studying dark matter distribution in the universe,
from which one can extract information on the basic cosmological parameters
(e.g., \citet{Massey2007, Kilbinger2013}).
Forthcoming weak-lensing surveys are aimed at measuring cosmic shear
over a wide area of more than 1000 ${\rm deg}^2$.
These observations will address important questions of dark matter and dark energy 
at unprecedented precision.

Unfortunately, major statistical methods to make the best use of WL data in upcoming surveys are 
still under debate.
The problem originates from the fact that cosmic shear follows non-Gaussian probability distribution 
due to non-linear gravitational growth \citep{Sato2009}.
In order to incorporate the non-linear features accurately into WL statistics,
cosmological $N$-body simulations have been extensively used. 
Previous numerical studies \citep{Hilbert2009,Sato2009,Sato2011} have already provided 
important guides for cosmological studies with WL statistics.
There still remain several possible factors to be examined. 
One of the uncertainties in such studies is the effects of \textit{baryonic physics}. 
Modeling baryonic effects on the WL statistics is difficult 
because of the overall complexities in galaxy formation.
Recent numerical simulations and semi-analytic methods 
successfully reproduce key observational data
\citep{Duffy2010,Martizzi2012,Schaye2014,Okamoto2014,
Martizzi2014,Schaller2014b,Schaller2014a,Velliscig2014,Pike2014}. 
Some of these studies focus on active galactic nuclei (AGN) feedback, 
which quenches star formation in massive halos and 
may solve \textit{over cooling problem} (i.e. the over production of 
stars in numerical simulations). 
These simulations also show that baryonic physics can change 
significantly the distribution 
of both dark matter and baryons within a halo.
However, since baryonic effects are expected to be weak at large scale, 
most of WL studies so far are based on simulations 
with dark matter component only.
For future `precision cosmology' with WL,
neglecting various astrophysical processes
may lead to undesirable bias of cosmological parameter estimation,
as has been pointed out by several studies
 \citep{Jing2006, Semboloni2011,Zentner2013,Yang2013, Mohammed2014}. 
Hence, it is crucial and timely to study the effects 
of baryonic physics on WL statistics in detail.

In this paper, we use several statistics to 
extract the non-Gaussian information of WL maps.
In addition to power spectrum (PS) of weak lensing convergence,  
we consider peak counts and the Minkowski Functionals (MFs).
\citet{Kratochvil2010} show that peak counts on lensing map 
can be indeed useful for cosmological parameter estimation. 
High-$\sigma$ lensing peaks are likely associated with massive 
dark matter halos along a line of sight
\citep{Hamana2004, Yang2011, Hamana2012}.
MFs are morphological statistics for a given multi-dimensional field.
MFs are one of the good measure of topology in WL and contain 
the suitable cosmological information 
beyond two-point statistics
\citep{Munshi2012,Kratochvil2012,Shirasaki2012,Shirasaki2014}.
Throughout this paper, we use the terms ``local statistics'' and
``non-local statistics''. 
Shear two-point correlation functions and the corresponding power spectrum 
are measured directly from shear that is obtained from local measurements,
i.e., from galaxy ellipticities.
Peak counts and MFs are {\it non-local} statistics, because
they can be obtained from convergence maps that are derived by
integrating shear over the region of interest.
In order to clarify the baryonic effects on the WL statistics,
we utilize a large set of hydrodynamical $N$-body simulations including various baryonic processes.
We then study how the baryonic effects bias cosmological parameter estimation with WL measurement.
The rest of this paper is organized as follows. 
In Section \ref{sec:statistics}, 
we summarize the basics of WL statistics of interest and
the implementations to estimate the statistics for a given WL data.
In Section \ref{sec:simulation}, 
we describe our simulation set that includes cosmological simulations
with baryonic physics.
In Section \ref{sec:analysis}, we provide the details of analysis performed in this paper.
In Section \ref{sec:results}, we show the impact of baryonic physics on WL statistics.
We also perform a Fisher analysis to present the expected cosmological constraints in upcoming lensing surveys.
We show the results of $\chi^2$ analysis to quantify the baryonic 
effects on parameter estimation with WL statistics.
Concluding remarks and discussions are given in Section \ref{sec:conclusion}.

\section{WEAK LENSING STATISTICS}
\label{sec:statistics}

We first summarize the basics of gravitational lensing by large-scale structure.
Weak gravitational lensing effect is characterized by
the image distortion of a source object by the 
following 2D matrix:
\beqa
A_{ij} = \frac{\partial \beta^{i}}{\partial \theta^{j}}
           \equiv \left(
\begin{array}{cc}
1-\kappa -\gamma_{1} & -\gamma_{2}  \\
-\gamma_{2} & 1-\kappa+\gamma_{1} \\
\end{array}
\right), \label{distortion_tensor}
\eeqa
where the observed position of a source object is denoted by $\bd{\theta}$, 
the true position is $\bd{\beta}$,
$\kappa$ is convergence,
and $\gamma$ is shear.
In weak field, where gravitational potential is small compared with $c^2$,
each component of $A_{ij}$ can be related to
the second derivative of the gravitational potential $\Phi$ as
\beqa
A_{ij} &=& \delta_{ij} - \Phi_{ij}, \label{eq:Aij} \\
\Phi_{ij}  &=&\frac{2}{c^2}\int _{0}^{\chi}{\rm d}\chi^{\prime} f(\chi,\chi^{\prime}) 
\frac{\partial^2}{\partial x_{i}\partial x_{j}}
\Phi[r(\chi^{\prime})\bd{\theta},\chi^{\prime}], \label{eq:shear_ten}\\
f(\chi,\chi^{\prime}) &=& \frac{r(\chi-\chi^{\prime})r(\chi^{\prime})}{r(\chi)},
\eeqa
where $r(\chi)$ is angular diameter distance, 
and $x_{i}=r\theta_{i}$ represents physical separation \citep[]{Bartelmann2001,Munshi2008}.
By using the Poisson equation and Born approximation
\citep[]{Bartelmann2001,Munshi2008}, 
one can express weak lensing convergence field as
\beqa
\kappa(\bd{\theta},\chi)= \frac{3}{2}\left(\frac{H_{0}}{c}\right)^2 \Omega_{\rm m}
\int _{0}^{\chi}{\rm d}\chi^{\prime} f(\chi,\chi^{\prime}) 
\frac{\delta[r(\chi^{\prime})\bd{\theta},\chi^{\prime}]}{a(\chi^{\prime})}. \label{eq:kappa_delta}
\eeqa
Born approximation yields sufficiently accurate two-point statistics \citep[e.g.,][]{Schneider1998}.
In the present paper, we take into account the non-linearity of convergence shown 
in Eq.~(\ref{eq:shear_ten})
by performing ray-tracing simulations through the matter density field obtained
from cosmological simulations.

\subsection{Observables}
Here, we summarize three different statistics of weak lensing convergence field,
power spectrum (PS), 
peak counts, and Minkowski Functionals (MFs).
PS has complete cosmological information 
only if statistical properties of matter fluctuation follows Gaussian distribution.
However, non-linear structure formation induced by gravity 
inevitably makes the fluctuation deviate significantly from Gaussian.
In order to extract cosmological information, 
we will also use non-local statistics, 
i.e. peak counts and MFs.

\subsubsection{Power spectrum}
PS is one of the basic statistics for modern cosmology.
For a convergence field $\kappa$, PS is defined by the two point correlation in Fourier space:
\begin{eqnarray}
\langle \tilde{\kappa}(\bm{\ell}) \tilde{\kappa}^*(\bm{\ell}^{\prime}) \rangle 
= (2\pi)^2 \delta_D (\bm{\ell}-\bm{\ell}^{\prime})P_\kappa (\ell),
\end{eqnarray}
where the multipole $\ell$ is related with angular scale through $\theta=\pi/\ell$. 
By using Limber approximation \citep{Limber:1954zz,Kaiser:1991qi}
and Eq.~\eqref{eq:kappa_delta}, 
one can derive the convergence power spectrum as 
\beqa
P_{\kappa}(\ell) &=& \int_{0}^{\chi_s} {\rm d}\chi \frac{W(\chi)^2}{r(\chi)^2} 
P_{\delta}\left(k=\frac{\ell}{r(\chi)},z(\chi)\right)
\label{eq:kappa_power},
\eeqa
where $P_{\delta}(k)$ represents the three dimensional matter power spectrum, 
$\chi_s$ is comoving distance to source galaxies 
and $W(\chi)$ is the lensing weight function defined as
\beqa
W(\chi) = \frac{3}{2}\left(\frac{H_{0}}{c}\right)^2 
\Omega_{\rm m}
\frac{r(\chi_s-\chi)r(\chi)}{r(\chi_s)}(1+z(\chi)).
\eeqa

We follow \citet{Sato2009} to estimate the convergence PS from numerical simulations.
We measure the binned power spectrum of convergence field 
by averaging the product of Fourier modes $|\tilde{\kappa}(\bd{\ell})|^2$.
We use 20 bins logarithmically spaced in the range of $\ell = 100$ to $10^{5}$.
For parameter estimation performed in section 4, we re-compute PS using 10 bins 
logarithmically spaced in the range of $\ell = 100$ to $2000$.

\subsubsection{Peak count}
\label{sec:peak}

Peaks in convergence maps can be a probe of massive halos \citep{Hamana2004,Hamana2012} 
and thus contain cosmological information \citep{Yang2011}. 

In practice, peaks on convergence map is defined by a local maxima on the ``smoothed" map.
We do the map-smoothing because the observed lensing field is significantly contaminated 
by the intrinsic ellipticities of source galaxies. 
The contaminant is called shape noise, which is indeed 
the major contribution to the measured shape of source galaxies. 
In practice, we use a Gaussian window function
in order to reduce the effect of shape noises on WL statistics.
The smoothed convergence $\mathcal{K}(\bm{\theta};\theta_G)$
can be written as convolution with a filter function of $W_{G}$:
\begin{eqnarray}
\mathcal{K}(\bm{\theta};\theta_G) 
= \int {\rm d}^2\phi \ W_G (\bm{\theta}-\bm{\phi};\theta_G) \kappa(\bm{\phi}),
\end{eqnarray}
where $\theta_G$ is the smoothing scale and $W_G$ is a gaussian filter given by
\begin{eqnarray}
W_G (\bm{\theta}) 
= \frac{1}{\pi \theta_G^2} \exp \left(-\frac{\theta_1^2+\theta_2^2}{\theta_G^2} \right). \label{eq:gauss_sm}
\end{eqnarray}

One can evaluate the smoothed convergence due to an isolated massive cluster at a given redshift
by assuming the universal matter density profile of dark matter halos \citep[e.g.,][]{Navarro:1996gj}.
A simple theoretical framework to predict the number density 
of the $\mathcal{K}$ peaks is presented by \citet{Hamana2004,Maturi2011}.
Their calculation yields reasonable results
when the signal-to-noise ratio of $\mathcal{K}$ due to massive halos is larger than $\sim4$
\citep[see,][for details]{Hamana2004}.
\citet{2010ApJ...719.1408F} also consider more detailed calculation 
by including the statistical properties of shape noise and the impact of shape noise 
on peak position.

In order to locate peaks on a discretized 
map obtained from numerical simulations,
we define the peak as a pixel that is higher than eight neighboring pixels.
We then measure the number density of peaks as a function of $\mathcal{K}$.
We exclude the region within 2$\theta_{G}$ from the boundary of the map 
in order to avoid the effect of incomplete smoothing.
In this paper, we divide peaks into two subgroups: 
``medium peaks (MPs)" and ``high peaks (HPs)".
We define MPs and HPs by the peak height
in a similar manner to \citet{Yang2013}.
The former is defined by the covergence peak with 
$1.0 \le \mathcal{K}_\mathrm{peak}/\sigma_\mathrm{noise} \le 3.0$
whereas the latter corresponds to the peak with  
$3.0 \le \mathcal{K}_\mathrm{peak}/\sigma_\mathrm{noise} \le 5.0$.
Here, $\sigma_\mathrm{noise}$ is the rms of shape noise on smoothed map 
given by Eq.~(\ref{eq:shape_noise_sm}).
It is important to note that 
these peaks are thought to have different physical origins.
MPs are likely caused by the shape noise or/and 
several dark matter halos aligned along a line of sight \citep{Yang2011},
whereas HPs are associated with individual massive dark matter halos \citep[e.g.,][]{Hamana2004}.
Throughout the paper, 
we set the number of bins to be 10 for parameter estimation when measuring peak count of each subgroup.

\subsubsection{Minkowski Functionals}
\label{sec:MFs}

MFs are morphological descriptors for smoothed random fields. 
There are three kinds of MFs for two-dimensional maps. 
Each MFs of $V_0$, $V_1$, and $V_2$ represent 
the area above the threshold $\nu$, 
the total boundary length,
the integral of geodesic curvature along the contours.
They are given by
\begin{eqnarray}
V_0(\nu) &\equiv& \frac{1}{A} \int_{Q_\nu} {\rm d}a, \\
V_1(\nu) &\equiv& \frac{1}{A} \int_{\partial Q_\nu} \frac{1}{4} {\rm d}\ell, \\
V_2(\nu) &\equiv& \frac{1}{A} \int_{\partial Q_\nu} \frac{1}{2\pi} K {\rm d}\ell,
\end{eqnarray}
where $K$ is the geodesic curvature of the contours, 
${\rm d}a$ and ${\rm d}\ell$ represent the area and length elements, 
and $A$ is the total area. 
$Q_\nu$ and $\partial Q_\nu$ are denoted to be excursion sets
and boundary sets for the smoothed field $\mathcal{K}(\bm{x})$, respectively.
They are defined by
\begin{eqnarray}
Q_\nu = \{ \bm{x}|\mathcal{K} (\bm{x}) > \nu \}, \\
\partial Q_\nu = \{ \bm{x}|\mathcal{K} (\bm{x}) = \nu \}.
\end{eqnarray}
In particular, 
$V_2$ is equivalent to 
a kind of genus statistics and equal to the number of connected regions above the threshold, 
minus ones below the threshold.
Therefore, for high thresholds, $V_2$ is essentially equivalent to the number of peaks.

For a two-dimensional Gaussian random field, 
the expectation values for MFs can be described by analytic functions 
as shown in \citep{1986PThPh..76..952T}:
\begin{eqnarray}
V_{0}(\nu)&=&\frac{1}{2}\left[1-{\rm erf}\left( \frac{\nu-\mu}{\sigma_{0}}\right)\right],
\label{eq:v0_gauss} \\
V_{1}(\nu)&=&\frac{1}{8\sqrt{2}}\frac{\sigma_1}{\sigma_0}\exp\left( -\frac{(\nu-\mu)^2}{\sigma_0^2}\right),
\label{eq:v1_gauss} \\
V_{2}(\nu)&=&\frac{\nu-\mu}{2(2\pi)^{3/2}}\frac{\sigma_{1}^2}{\sigma_{0}^3}\exp \left( -\frac{(\nu-\mu)^2}{\sigma_0^2} \right),
\label{eq:v2_gauss}
\end{eqnarray}
where $\mu=\langle \mathcal{K} \rangle$, $\sigma_{0}^2 = \langle \mathcal{K}^2 \rangle - \mu^2$, and
$\sigma_{1}^2 = \langle |\nabla \mathcal{K}|^2 \rangle$.
Although MFs can be evaluated perturbatively 
if the non-Gaussianity of the field is weak \citep{Matsubara2003,Matsubara2010},
it is difficult to evaluate MFs of highly non-Gaussian field \citep{Petri2013}.
In this paper, we pay a spatial attention to the non-Gaussian cosmological information 
obtained from convergence MFs.

For discretized $\mathcal{K}$ maps, 
we employ following estimators, as shown in, e.g., \citet{Kratochvil2012},
\begin{eqnarray}
V_0(\nu) &=& \frac{1}{A} \int \Theta (\mathcal{K}-\nu){\rm d}x{\rm d}y, \\
V_1(\nu) &=& \frac{1}{4A} \int \delta (\mathcal{K}-\nu) \sqrt{\mathcal{K}_x^2+\mathcal{K}_y^2} 
{\rm d}x{\rm d}y, \\
V_2(\nu) &=& \frac{1}{2\pi A} \int \delta (\mathcal{K}-\nu) \frac{2\mathcal{K}_x\mathcal{K}_y\mathcal{K}_{xy}-\mathcal{K}_x^2\mathcal{K}_{yy}-\mathcal{K}_y^2\mathcal{K}_{xx}}{\mathcal{K}_x^2+\mathcal{K}_y^2} {\rm d}x{\rm d}y,\nonumber \\
\end{eqnarray}
where $\Theta(x)$ is the Heaviside step function and 
$\delta(x)$ is the Dirac delta function. 
The subscripts represent differentiation with respect to $x$ or $y$. 
The first and second differentiation are evaluated with finite difference.
We precompute MFs for 100 equally spaced bins of 
$\nu'=(\nu-\langle \mathcal{K} \rangle)/\sigma_0$ between $-10$ to $10$.
For cosmological parameter estimation, we recalculate values on
equally spaced 10 bins in the range $-3.0 \leq \nu' \leq 3.0$ from 100 bins.

\section{SIMULATION}
\label{sec:simulation}

\subsection{$N$-body simulations}
\begin{table*}
\centering
\caption{Cosmological parameters used for simulations.\label{tb:params}}
\begin{tabular}{@{}lcccccccl}
\hline
\hline
Run & $w$ & $10^9A_s$ & $\Omega_\textrm{m}$ & $\Omega_\Lambda$ & $\sigma_8$ & No. of sim. & No. of maps & Explanation \\ \hline
DM & $-1.0$ & 2.41 & 0.279 & 0.721 & 0.821 & 10 & 100 &CDM only fiducial model\\
BA & $-1.0$ & 2.41 & 0.279 & 0.721 & 0.821 & 10 & 100 & CDM and adiabatic gas \\
FE & $-1.0$ & 2.41 & 0.279 & 0.721 & 0.821 & 10 & 100 & CDM and baryonic processes \\
High $\Omega_\textrm{m}$ & $-1.0$ & 2.41 & 0.302 & 0.698 & 0.872 & 10 & 100 & $1\sigma$ higher $\Omega_\textrm{m}$ model \\
Low $\Omega_\textrm{m}$ & $-1.0$ & 2.41 & 0.256 & 0.744 & 0.767 & 10 & 100 & $1\sigma$ lower $\Omega_\textrm{m}$ model  \\
High $w$ & $-0.8$ & 2.41 & 0.279 & 0.721 & 0.766 & 10 & 100 & higher $w$ model \\
Low $w$ & $-1.2$ & 2.41 & 0.279 & 0.721 & 0.860 & 10 & 100 & lower $w$ model \\
High $A_s$ & $-1.0$ & 2.51 & 0.279 & 0.721 & 0.838 & 10 & 100 & $1\sigma$ higher $A_s$ model\\
Low $A_s$ & $-1.0$ & 2.31 & 0.279 & 0.721 & 0.804 & 10 & 100 & $1\sigma$ lower $A_s$ model\\
\hline
\end{tabular}
\tablecomments{
	Two parameters ($\Omega_\textrm{m}$, $10^9A_s$) 
	are varied by 1$\sigma$ value of {\it WMAP} nine-year result \citep{Hinshaw2013} 
	and we also adjust $\Omega_\Lambda$ accordingly to keep the universe spatially flat. 
	We also show the normalization of linear matter power spectrum $\sigma_8$, which is a derived parameter.
	}
\end{table*}

We are interested in non-linear gravitational evolution
of large-scale structure.
In order to follow the evolution accurately, 
we run cosmological $N$-body simulations. 
We use parallelized tree-PM code {\tt Gadget-3} \citep{Springel2005} 
with baryonic processes (discussed below) 
to follow structure formation from an early epoch ($z=99$) to present ($z=0$). 
The initial conditions are generated by {\tt MUSIC} code \citep{Hahn2011}, 
which is based on the second order Lagrangian perturbation theory \citep[e.g.,][]{2006MNRAS.373..369C}. 
The transfer function is generated by the linear Boltzmann code {\tt CAMB} \citep{Lewis2000}.
The volume of each simulation box is comoving $240\ \mathrm{Mpc}/h$ on a side. 
We adopt the fiducial cosmological parameters as follows: 
matter density $\Omega_\mathrm{m}=0.279$, 
baryon density $\Omega_\mathrm{b} =0.0463$, 
dark energy density $\Omega_\Lambda=0.721$, 
Hubble parameter $h=0.70$, 
spectral index $n_s = 0.972$ and 
amplitude of scalar perturbation $A_s=2.41\times 10^{-9}$ at the pivot scale $k=0.002\ \mathrm{Mpc}^{-1}$. 
These parameters are consistent with 9-year 
\textit{Wilkinson Microwave Anisotropy Probe} (\textit{WMAP}) result \citep{Hinshaw2013}.

In this paper, we perform three kinds of cosmological simulations; 
cold dark matter (CDM) simulation (denoted as ``DM") and two baryonic simulations.
In order to model the degeneracy between cosmological parameters in WL statistics, 
we run six CDM only simulations with one cosmological parameter varied.
Cosmological parameters for our models are summarized in Table \ref{tb:params}.
The number of particles is set to be $512^3$ for CDM only simulations 
and $2\times512^3$ for baryonic simulations, 
which consist of both CDM and gas particles. 
For the fiducial cosmological model, the mass of a particle is found to be 
$m_p=7.97\times 10^9 M_{\odot}/h$ and $m_\mathrm{cdm}=6.65\times 10^9 M_{\odot}/h, m_\mathrm{gas}=1.32\times 10^9 M_{\odot}/h$ for baryonic simulations. 
Our baryonic simulations are based on two models denoted as BA and FE. 
BA model contains adiabatic gas particles, 
which exert only adiabatic pressure in a Smoothed Particle Hydrodynamics (SPH) manner. 
In FE model, we employ the galaxy formation `recipes' of \citet{Okamoto2014}. 
Our FE model corresponds to SN+AGN model in their paper with some modifications, 
which include star formation, radiative cooling, supernova (SN) feedback, 
stellar wind feedback and an {\it ad hoc} active galactic nuclei (AGN) feedback.
This prescription saves computation time and contains less free parameters
than solving fully the evolution of black holes.
In their model, the radiative cooling rate is exponentially suppressed
when the velocity dispersion within a dark matter halo exceeds some threshold. 
Our simulations employ a simpler method that 
switches off radiative cooling when the local velocity dispersion reaches the threshold. 
\citet{Okamoto2014} report that the sudden change of the cooling function 
leads to somewhat artificial increase of the stellar mass,
but we expect this modification does not affect the final results significantly.

\begin{figure}
\centering \includegraphics[clip, width=0.45\textwidth]{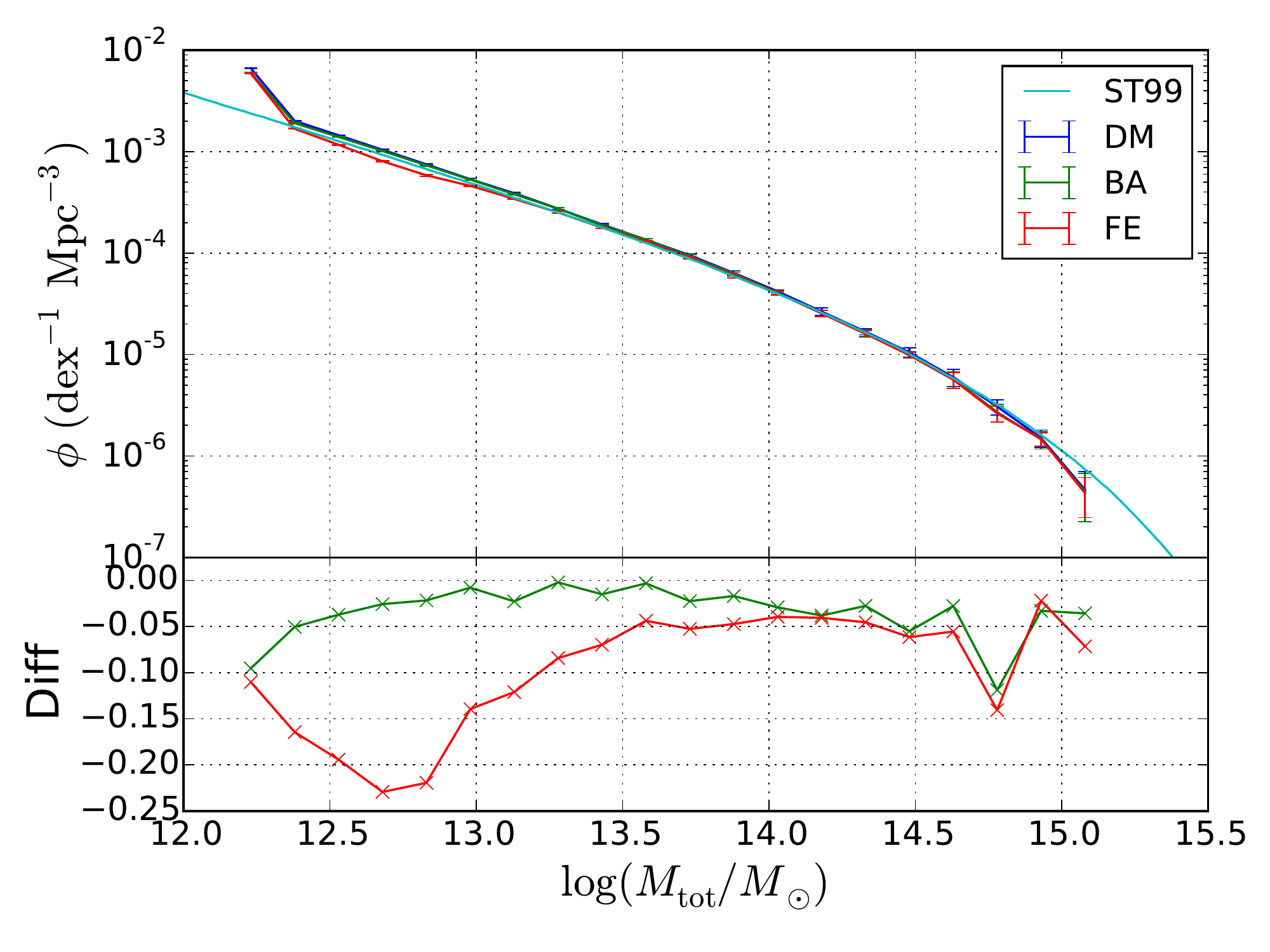}
\caption{
	{\it Top panel}: We plot the halo mass functions at $z=0$ measured from our set of simulations. 
	The horizontal axis represents total mass of halos which include dark matter, 
	gas and star particles. 
	The blue, green and red line corresponds to simulation results 
        averaged over 10 realizations 
	of DM model, BA model and FE model, respectively. 
	The error bars represent the standard deviation of the ten realizations 
        for each model.
	The cyan line shows model prediction by \citet{Sheth1999}.
	{\it Bottom panel}: The fractional difference of baryonic models
         from the fiducial model.
	}
\label{fig:mf}
\end{figure}

\begin{figure}
\centering \includegraphics[clip, width=0.45\textwidth]{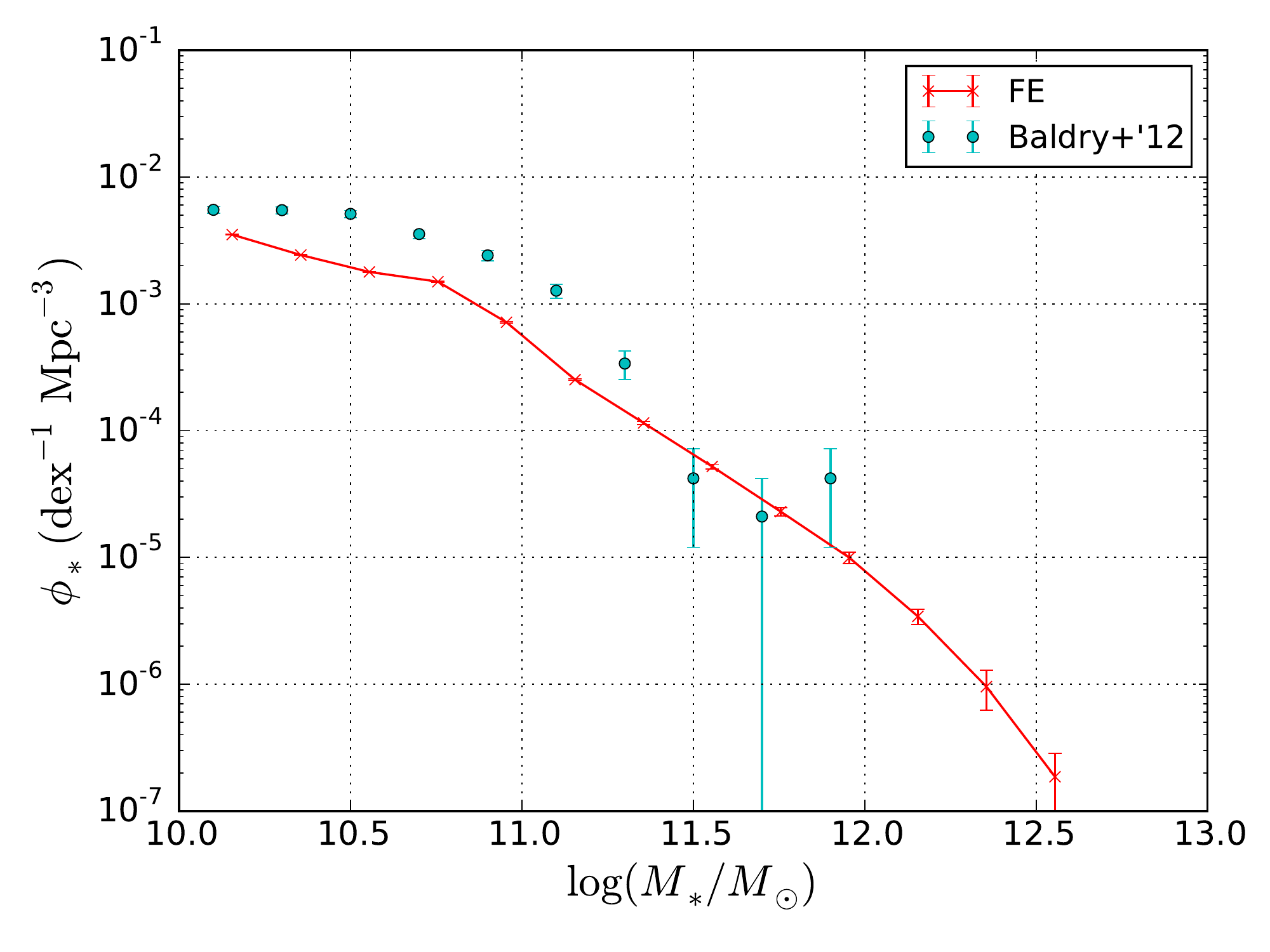}
\caption{
	Galaxy stellar mass functions at $z=0$. 
	The cyan points show observational estimates from \citet{Baldry2012}. 
	The red line illustrates the result from FE simulation averaged over 10 realizations
        and  
	the error bars represent the standard deviation.
	}
\label{fig:smf}
\end{figure}
In order to check basic statistics of our simulations,
we measure halo mass function for each model 
and stellar mass function for FE model.
We run a friends-of-friends (FoF) halo finder {\tt SubFind} 
\citep{Springel2001, Dolag2009} to identify halos in simulations.
For our simulations with baryons, the procedure is different 
from that for CDM only simulation.
First, using only dark matter particles, 
we find FoF groups which contain more than 31 particles. 
Next, we find gas particles for BA model and gas and star particles for FE model
linked with a dark matter particle which belongs to a FoF group. 
Finally, we separate each FoF group into a central halo and substructures. 
In FE model, the stellar component of a halo is called `galaxy'
and `halo' includes the baryonic components and dark matter.
Figure~\ref{fig:mf} and \ref{fig:smf} represent halo and stellar mass 
functions measured from simulations at $z=0$. 
We compare the simulation result with theoretical 
prediction of \citet{Sheth1999}.
Figure~\ref{fig:mf} clearly shows our simulation results
are well described by the theoretical prediction over the typical cluster mass scale, 
while there are appreciable discrepancies 
between DM and FE simulations in the halo mass of less than 
$\sim10^{13.5} M_{\odot}$.
This is partly induced by SN explosions and/or stellar winds.
Energy released by SN explosions or stellar winds expel gas particles from the halo 
but massive halos can retain the particles by the deep gravitational potential
well. As a result, the mass function for our FE model is slightly smaller at small masses.
We also compare the stellar mass functions of our simulation results and the observation 
of galaxy stellar mass function in \citet{Baldry2012}.
The stellar mass distribution in our FE model is in reasonable agreement
with the observation at $z\sim0$.
Because we adopt a simpler model than the original
implementation of \citet{Okamoto2014} that includes radiation pressure feedback,
our FE run produces less small-mass galaxies.
Also, our simulation results are inaccurate at stellar masses less than $\sim 10^{10} M_{\odot}$,
because such small galaxies contain only a few star particles.
These small discrepancies are, however,
unimportant in the present paper because we draw
our main conclusions through comparisons of multiple
sets of simulations with and without baryonic
components, rather than through comparisons
of different feedback models.

\subsection{Ray-tracing simulations}

\begin{figure*}
\centering \includegraphics[clip, width=0.95\textwidth]{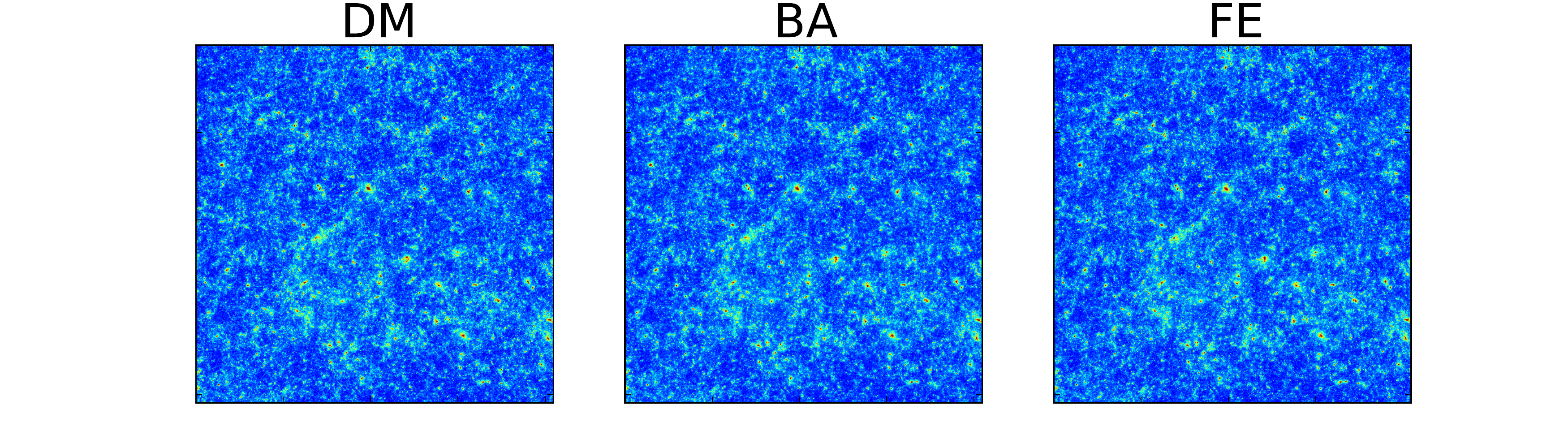}
\caption{We show sample convergence maps of DM, BA and FE simulations 
without noise and smoothing. Red or blue area corresponds to high or low convergence respectively. There are many red points, which are identified as peaks.}
\label{fig:map}
\end{figure*}

For ray-tracing simulations of gravitational lensing, 
we utilize multiple simulation boxes to generate light-cone outputs 
similarly to \citet{White2000}, \citet{Hamana2001}, 
and \citet{Sato2009}.
Details of the configuration are found in the last reference.

We place the simulation outputs to fill the past light-cone of a hypothetical observer 
with an angular extent $5^{\circ}\times 5^{\circ}$, from $z=0$ to 1.
The angular grid size of our maps is set to be $5^{\circ}/4096\sim 0.073$ arcmin.
We randomly rotate and shift the simulation boxes 
in order to avoid the same structure appearing 
multiple times along a line-of-sight.
In total, we generate 100 independent lensing maps 
for the source redshift of $z_{\rm source}=1$
\footnote{For $z_{\rm source} = 1$, the lensing weight function $W(\chi)$ has a
maximum at $z \sim 0.5$. This means that our lensing statistics
are not sensitive to the large scale structure above $z=0.7-0.8$.}.
We show an example of convergence maps obtained 
from each baryonic model (DM, BA, and FE) in Figure~\ref{fig:map}. 
Note that each realization of three models 
use the same random seed when we generate initial conditions and multiple planes.

In order to make the mock lensing maps more realistic, 
we add random gaussian noises as shape noise 
to the simulated convergence data (e.g., \citet{Kratochvil2010} and \citet{Shirasaki2012}):
\begin{eqnarray}
\langle \kappa_\mathrm{noise}(x_1,x_2) \kappa_\mathrm{noise}(x_1',x_2')\rangle 
= \frac{\sigma_\gamma^2}{n_\mathrm{gal}A_\mathrm{pix}} \delta_{x_1,x_1'}\delta_{x_2,x_2'},\nonumber \\
\end{eqnarray}
where $\delta_{x, y}$ is the Kronecker delta symbol, 
$n_\mathrm{gal}$ is the number density of source galaxies, 
$\sigma_\gamma$ is the rms of shape noise 
and $A_\mathrm{pix}$ is the solid angle of a pixel.
In the following, we adopt $\sigma_\gamma = 0.4, n_\mathrm{gal}=30\ \mathrm{arcmin}^{-2}$.
These values are expected to be typical for HSC survey. 

As described in Section~\ref{sec:peak} and \ref{sec:MFs},
smoothing is required to measure non-local statistics of convergence from noisy WL data.
For a given smoothing scale $\theta_{G}$,
the rms of the shape noise after gaussian smoothing is given by \citet{VanWaerbeke2000}
\begin{eqnarray}
\sigma_\mathrm{noise} = 0.0291
\left(\frac{\sigma_{\gamma}}{0.4}\right)
\left(\frac{\theta_G}{1\, {\rm arcmin}}\right)^{-1} 
\left(\frac{n_\mathrm{gal}}{30\, {\rm arcmin}^{-2}}\right)^{-1/2}.\nonumber \\
\label{eq:shape_noise_sm}
\end{eqnarray}

\section{ANALYSIS}
\label{sec:analysis}

We quantify the effects of the baryonic processes 
on weak lensing analyses in terms of errors and bias in 
cosmological parameter estimation.
We consider primarily a lensing survey with a sky coverage of 1400 deg$^2$,
i.e., the ongoing wide-field survey by Subaru Hyper Suprime-Cam (HSC).
In the following, we describe in detail
the calculation of the ensemble averages of three statistics
and the covariance matrix 
in order to derive statistical implications and to estimate 
cosmological parameters 
$\bm{p} = (\Omega_\textrm{m},w, 10^9A_s)$. 

\subsection{Theoretical model and covariance of lensing statistics}
\begin{table}
\caption{The configuration of bins for each observable\label{tb:bin}}
\centering
\begin{tabular}{@{}lcccl}
\hline
\hline
Observable & Range & No. of bins \\ \hline
PS & $100 \le \ell \le 2000$ & 10 \\
MP & $1.0\le \mathcal{K}_\mathrm{peak}/\sigma_\mathrm{noise} \le 3.0$ & 10  \\
HP & $3.0\le \mathcal{K}_\mathrm{peak}/\sigma_\mathrm{noise} \le 5.0$ & 10 \\
MFs & $-3.0 \le (\nu-\langle \mathcal{K} \rangle)/\sigma_0 \le 3.0 $& 10  for each \\
\hline
\end{tabular}
\tablecomments{We summarize the configurations of our statistical analysis.
Bins are linearly separated except for PS that are evaluated 
using logarithmically spaced bins.}
\end{table}

First, we calculate the theoretical template and covariance of the WL statistics
using our ray-tracing simulations.
The configuration of bins are given in Table~\ref{tb:bin}. 
The model template and covariance are based on our DM models.
Later, we examine the baryonic effects on WL cosmological analysis
by comparison with DM model and two baryonic models.

We represent the data vector as $N_i$
and denote the dimension of the data vector as $n$. 
Note that $n$ can be larger than ten, when multiple observables are combined.
We derive theoretical prediction of lensing statistics by averaging over $R=100$ realizations:
\begin{eqnarray}
\langle N_i (\bm{p}) \rangle \simeq \bar{N_i}(\bm{p})  \equiv \frac{1}{R} \sum_{r=1}^R N_i (r,\bm{p}).
\end{eqnarray}
In order to consider the cosmological parameter dependence on $N_{i}$, 
we employ linear interpolation based on DM models with seven different cosmologies.
Thus, we obtain the theoretical model of $N_{i}$ for a given cosmological model as follows:
\begin{eqnarray}
\bar{N_i}(\bm{p}) = \bar{N_i}(\bm{p}^0) + \sum_\alpha \frac{\bar{N_i}(\bm{p}^{\alpha+}_\alpha)-\bar{N_i}(\bm{p}^{\alpha-}_\alpha)}{p^{\alpha+}_\alpha-p^{\alpha-}_\alpha} (p_\alpha-p^0_\alpha),\nonumber \\
\end{eqnarray}
where $\alpha$ runs 1 to 3 and the superscript 0 means fiducial value 
$\bm{p}^0=(0.279,-1.0,2.41)$. 
$\bm{p}^{\alpha\pm}$ represents a vector with one parameter 
with a higher or lower value; 
for example $\bm{p}^{1+} = (0.302,-1.0,2.41)$ 
(the other parameter values are given in Table.\ref{tb:params}).

The covariance matrix of the data vector $N_{i}$ on 25 deg$^2$ maps 
is estimated as
\begin{eqnarray}
C_{ij} (\bm{p}) 
= \frac{1}{R-1} \sum_{r=1}^R [N_i (r,\bm{p})-\bar{N_i}(\bm{p})][N_j (r,\bm{p})-\bar{N_j}(\bm{p})].\nonumber \\
\label{eq:cov_25}
\end{eqnarray}
We ignore the cosmological dependence of $C(\bm{p})$
\citep[see, e.g.,][]{Eifler2009} 
and hence evaluate the covariance matrix by using 
the fiducial model, i.e. $C(\bm{p}^0)$.
For the fiducial HSC survey,
we simply scale the covariance matrix by survey area,
by multiplying the covariance matrix 
$C_{ij} (\bm{p}^0)$ in Eq.~(\ref{eq:cov_25}) by a factor of $25/1400$. 
When calculating the inverse covariance, 
we include a debiasing correction, 
the so-called Anderson-Hartlap factor $\alpha=(R-n-2)/(R-1)$ \citep{Hartlap2007},
where $R$ is the number of realizations and $n$ is the dimension of the data vector.

\subsection{Fisher analysis}
\label{fisher}

Fisher analysis gives a simple forecast for statistical confidence level 
of three cosmological parameters ($\Omega_\mathrm{m},w,10^9A_s$)
with WL statistics. 
The Fisher matrix is given by
\begin{eqnarray}
F_{ij} = \frac{1}{2} \mathrm{Tr} [A_i A_j + C^{-1}M_{ij}], \label{eq:fisher_matrix}
\end{eqnarray}
where $A_i = C^{-1} \partial C/\partial p_i$, 
$M_{ij} = 2 (\partial \bm{N}/ \partial p_i) (\partial \bm{N}/ \partial p_j)^T $. 
The first term vanishes 
when the cosmological dependence is weak \citep{Eifler2009}. 
To compute the second term, 
the first derivative is evaluated by the first order finite difference, 
which is given by
\begin{eqnarray}
\frac{\partial \bm{N}}{ \partial p_i} 
\simeq \frac{\bm{\bar{N}}(\bm{p}^{i+}_i)-\bm{\bar{N}}(\bm{p}^{i-}_i)}{p^{i+}_i-p^{i-}_i}.
\end{eqnarray}
Then, 
the marginalized error over the other two parameters is given by
\begin{eqnarray}
\sigma_i = \sqrt{(F^{-1})_{ii}}. \label{eq:onesig_fisher}
\end{eqnarray}

\subsection{Mock lensing surveys}

In order to make a realistic forecast in upcoming HSC survey, 
we employ the bootstrap method as in \citet{Yang2013}. 
Since our suite of WL maps consist of one hundred 25 deg$^2$ maps, 
we randomly choose $1400/25=56$ realizations from one hundred 25 deg$^2$ WL maps. 
Repeating this procedure one thousand times, 
we can get one thousand `mock' HSC WL maps.
The resulting maps are based on the fiducial cosmological model 
but we also have the same set from simulations with the different baryonic effects. 
Lensing statistics of interest (i.e. PS, peak counts and MFs) in HSC surveys 
are evaluated by averaging each statistics on a 25 deg$^2$ map over 56 realizations. 

We then utilize the HSC observables derived in this way
to investigate the impact of baryonic effects on parameter estimation. 
We perform $\chi^2$ minimization to $N_{i, {\rm HSC}}$ as follows:
\begin{eqnarray}
\chi^2 (r,\bm{p},m) 
= \sum_{i,j} \varDelta N_i (C^{-1})_{ij} \varDelta N_j ,\label{eq:chi2}
\end{eqnarray}
where $\varDelta N_i \equiv N_{i, {\rm HSC}}^m(r)-\bar{N}_i(\bm{p})$,
$r$ is the index of realizations and $1\leq r \leq 1000$,
$\bar{N}_i$ represents the theoretical model of $N_{i}$ for our dark matter only
(DM) model, and
$m$ represents the difference of model of baryonic physics, i.e., BA and FE.
Suppose we calculate the $i$-th data 
$N_{i, {\rm HSC}}$ from a dark matter only convergence map ($m=\mathrm{DM}$),
we should then find the resulting best-fit points distribute around the fiducial point 
$\bm{p}^0=(0.279,-1.0,2.41)$.
However, when one considers the case of $m= \mathrm{BA\ or\ FE}$, 
the center of the distribution of best-fit values could be biased in parameter space of $\bm{p}$
if baryonic effects induce discrepancies between $N_{i, {\rm HSC}}$ and $\bar{N}_i(\bm{p}^0)$.
That means $\chi^2$ values calculated from Eq.~(\ref{eq:chi2}) do not follow $\chi^2$ distribution
for baryonic models because the ensemble mean and the covariance matrix are
computed from the fiducial model, not baryonic models.

There is another way to estimate the bias of parameter estimation.
The parameter biases of the BA and FE models 
from fiducial parameters can be computed in the following manner \citep[see, e.g.,][]{Huterer2006}:
\begin{eqnarray}
\delta p_\alpha = 
\sum_{\beta} F^{-1}_{\alpha\beta} \sum_{i,j}
[\bar{N}_{i}^m-\bar{N}_i(\bm{p}^0)] 
(C^{-1})_{ij} \frac{\partial N_j}{\partial p_\beta},\label{eq:param_bias}
\end{eqnarray}
where $F$ is the fisher matrix given by Eq.~(\ref{eq:fisher_matrix})
and $\bar{N}_{i}^m$ represents the average of $N_{i}$ over 100 convergence maps 
for $m= \mathrm{BA\ and\ FE}$.

In the following, we quantify the baryonic effects on parameter estimation with WL surveys
by considering the distribution of $\chi^2$ in Eq.~(\ref{eq:chi2}) and
$\delta p_\alpha$ in Eq.~(\ref{eq:param_bias}) for our BA and FE models.
We then compare the bias of parameter estimations 
with the marginalized error given in Eq.~(\ref{eq:onesig_fisher}).


\section{RESULTS}
\label{sec:results}

\subsection{Baryonic effect on convergence statistics}

\begin{figure}
\centering \includegraphics[clip, width=0.45\textwidth]{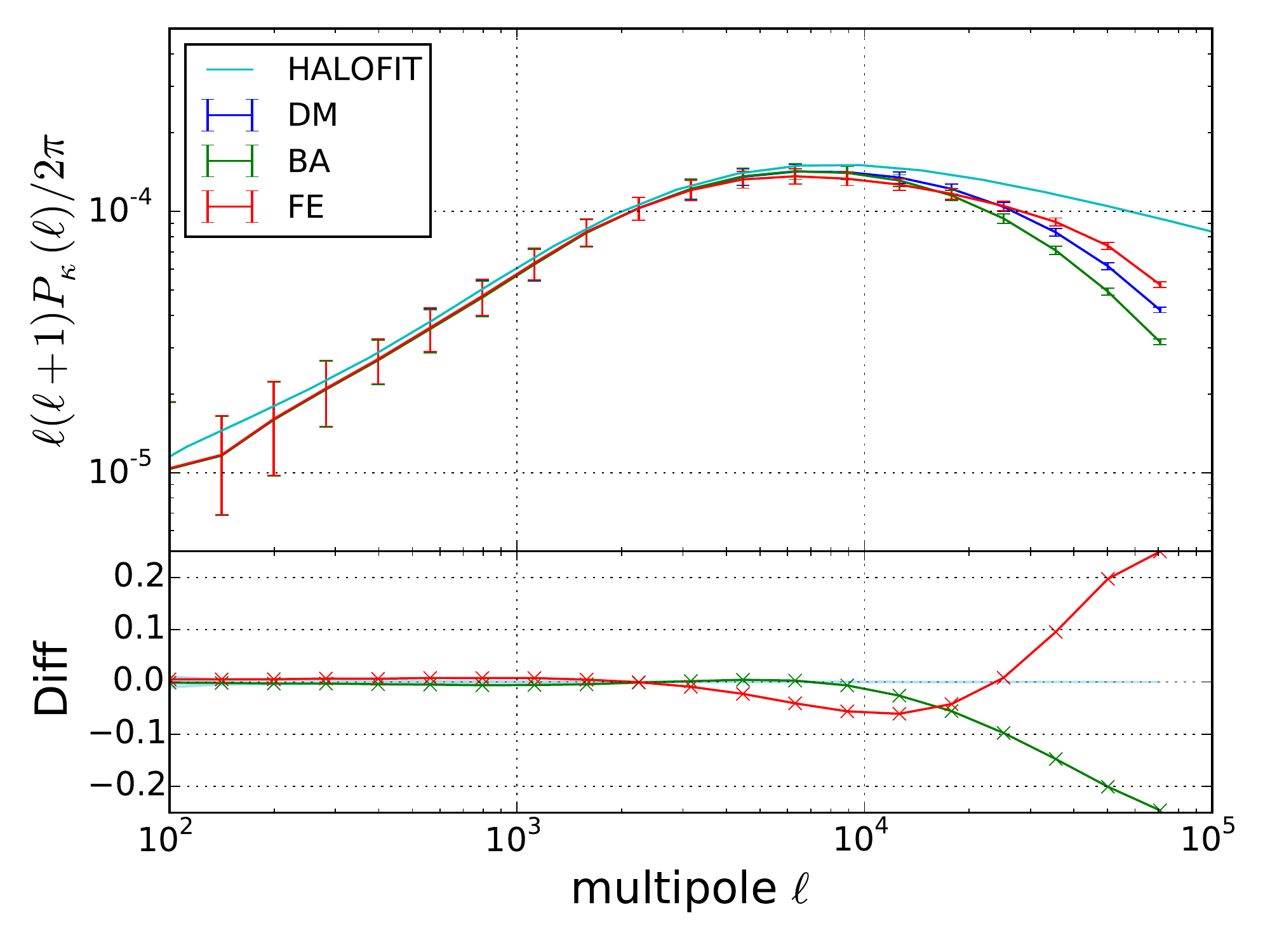}
\centering \includegraphics[clip, width=0.45\textwidth]{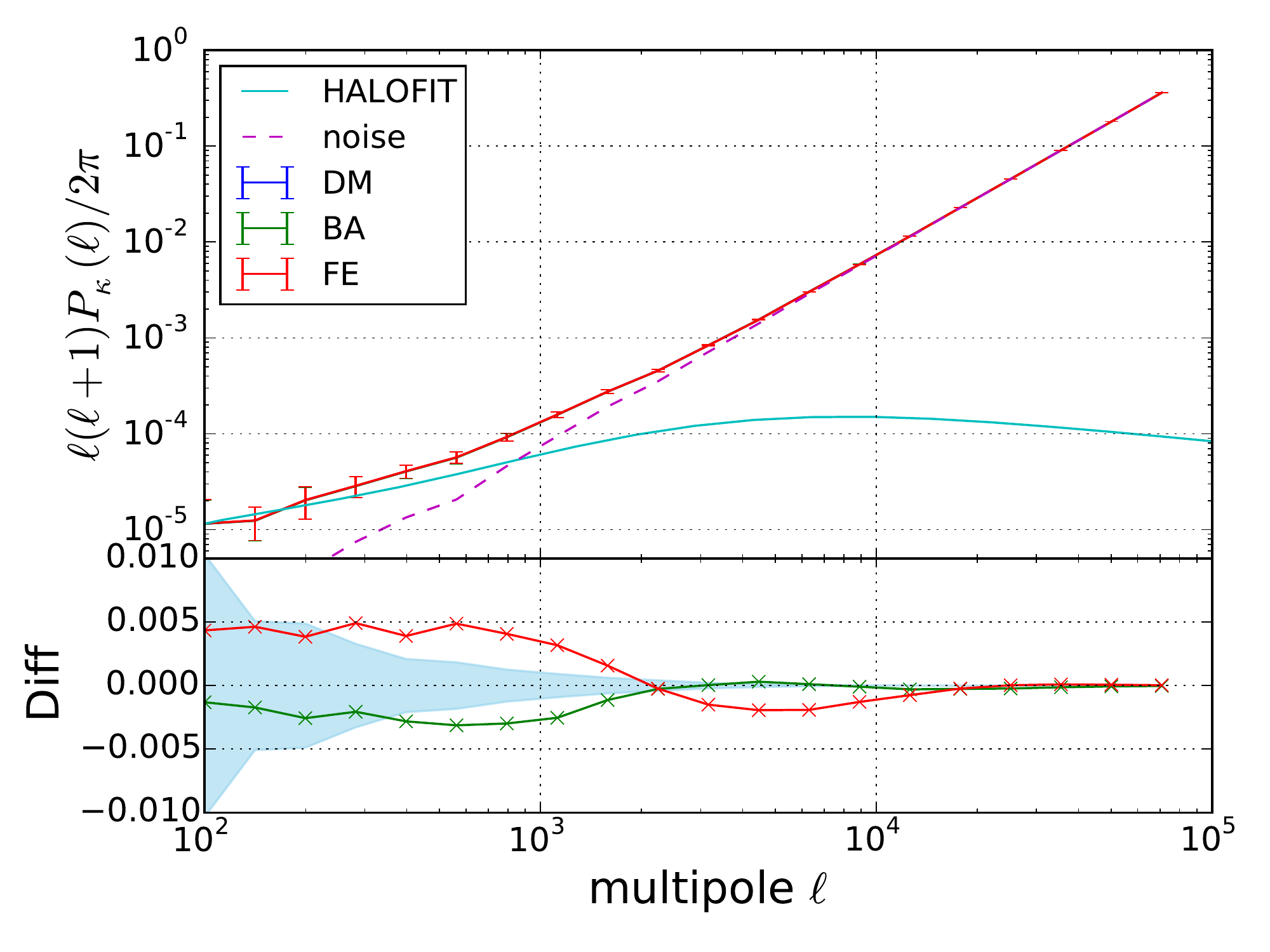}
\caption{We plot the convergence power spectrum from one hundred convergence maps. 
The top and bottom panel show the result without and with noise and smoothing. 
The solid cyan line shows the model prediction using {\it HaloFit} 
of \citet{Takahashi2012}.
The measured powers from simulations are slightly smaller than
the result of {\it HaloFit} due to finite box effect.
The dashed magenta line in the noisy case shows the
power spectrum of the noise. In the lower portion of each panel, we plot 
the fractional differences from the fiducial model and
the blue filled region corresponds to the expected statistical error of DM model
when one measures the power spectrum from one hundred 1400 deg$^2$ maps.
Note that this region does not correspond to the error bar, i.e. the standard deviation,
shown in the upper panel that is measured from our sample of 25 deg$^2$ maps.
And in the top panel, this expected error of power spectrum from noise free maps
is so small that it does not appear clearly.}
\label{fig:spec}
\end{figure}

We summarize the main results of 
the baryonic effects on WL statistics  in 
Figure~\ref{fig:spec}, \ref{fig:peak} and \ref{fig:minkowski}.

Figure~\ref{fig:spec} show PS from our maps, where 
DM, BA and FE models are indicated by blue, green and red lines respectively.
We also show the theoretical prediction 
by the cyan line which is calculated by Eq.~(\ref{eq:kappa_power}) 
with the modeling of matter power spectrum,
essentially an modified HaloFit, of \citet{Takahashi2012}
We plot the shot noise contribution in the bottom panel as the dashed magenta line.
The error bars represent the standard deviations over one hundred maps. 
The top and bottom panel represents to results from the maps with and without noises,
respectively. The lower portion in each panel shows the fractional differences of 
BA and FE models with respect to the DM model.

\subsubsection*{Power spectrum}

Compared with DM model,
gas pressure suppresses small-scale ($\ell>10000$)
structures in BA model.
On the other hand, in FE model feedback processes are efficient at the scales of $2000 < \ell < 10000$ but
cooling processes enhance the convergence power at the smaller scale $\ell > 20000$.
These features by cooling and feedback processes
are consistently found by \citet{Semboloni2011, Mohammed2014}.
However, the strength of the baryonic enhancement and suppression is different,
because of the details of baryonic physics implementations
and also partially because of the difference of cosmological parameters.

\subsubsection*{Peak count}
\begin{figure}
\centering \includegraphics[clip, width=0.45\textwidth]{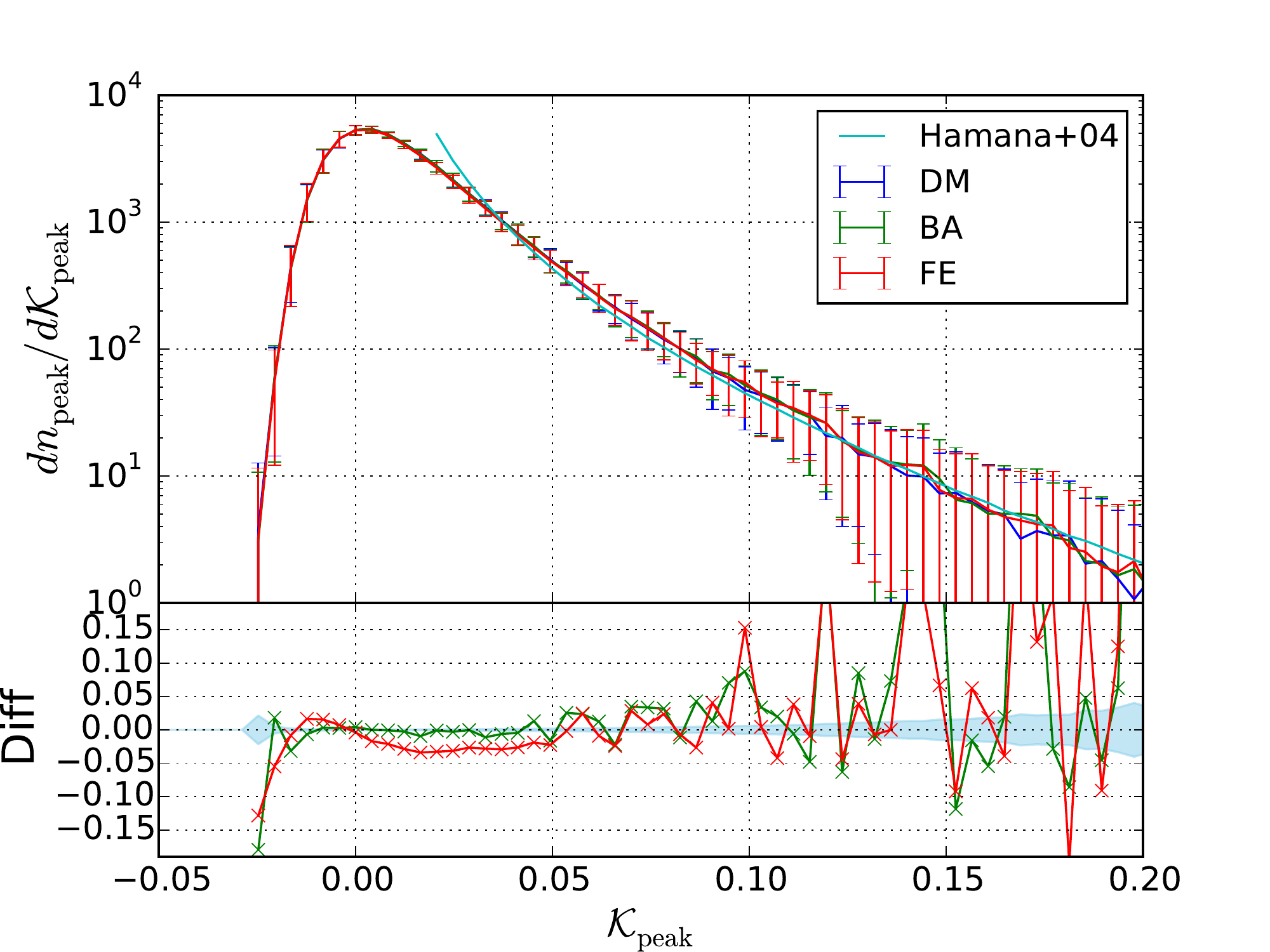}
\centering \includegraphics[clip, width=0.45\textwidth]{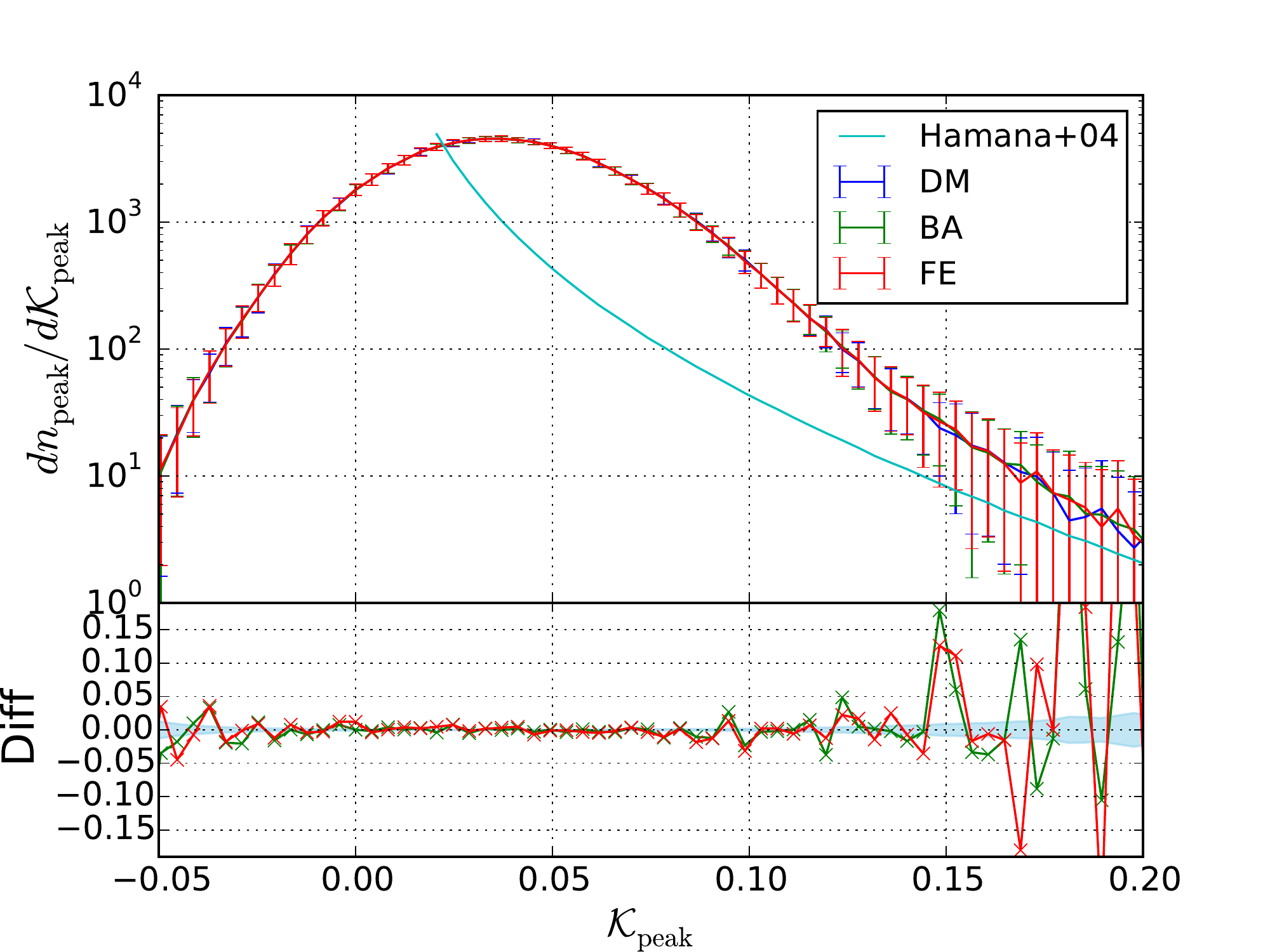}
\caption{The number of peaks per unit square degree. 
The top and bottom panel shows the result without and with noise. The solid cyan line shows the model prediction of peaks by \citet{Hamana2004}, which assumes that the halo mass function is described by \citet{Sheth1999} and the density profile of halos follow Navarro-Frenk-White profile \citep{Navarro1996,Navarro1997}. 
The lower portion in each panel shows the 
fractional difference from the fiducial model. 
The blue filled region represents expected Poisson error of DM model
from one hundred 1400 deg$^2$ maps.}
\label{fig:peak}
\end{figure}

Figure~\ref{fig:peak} shows the peak counts measured from 
the simulated convergence maps for the three models. 
We use the same combination
of line colors as in Figure ~\ref{fig:spec}.
The error bars represent standard deviations from one hundred maps.
The lower portion in each panel shows the fractional difference of
BA and FE models with respect to DM model, and the shaded region indicates
the Poisson error of the fiducial model, i.e. the square root of the number of peaks within a bin,
In this figure, we employ 5 bins per unit S/N ratio $\nu/\sigma_\mathrm{noise}$,
the peak height of convergence $\nu$
divided by $\sigma_\mathrm{noise}$.
For the analytical prediction,
we adopt  the model of \citet{Hamana2004}
with Sheth-Tormen \citep{Sheth1999} mass function
and Navarro-Frenk-White dark matter density profile \citep{Navarro1996,Navarro1997}.
We assume the relation between the halo mass and concentration parameter as in
\citep{Hamana2012},
\beqa
c(M,z) = 7.26 \left( \frac{M}{10^{12}M_{\odot} /h} \right)^{-0.086} (1+z)^{-0.71},
\eeqa
In Figure~\ref{fig:peak},
medium peaks (MPs) correspond to the peaks with
$ 0.03 \simlt \mathcal{K}_{\rm peak} \simlt 0.09 $,
while high peaks (HPs) are those with $ 0.09 \simlt \mathcal{K}_{\rm peak} \simlt 0.15 $.
HPs are typically associated with massive halos with the mass of $\sim10^{14}M_\odot$.
The number of HPs is less affected by baryonic effects
because the baryonic processes do not cause very strong effect to change 
the number of such massive halos,
as seen in Figure~\ref{fig:mf}.
MPs often originate from multiple
halos with masses of $10^{12}-10^{13} M_{\odot}$
aligned in the line of sight direction.
Baryonic physics, such as SN and AGN feedback,
reduce the masses of these halos (see Section~\ref{sec:peak}).
Hence the baryonic effects would decrease the number of MPs,
but the difference is largely made unimportant in the noisy convergence maps
because MPs are affected significantly by shape noise \citep{Yang2011}.

\subsubsection*{Minkowski Functionals}
\begin{figure*}
\centering \includegraphics[clip, width=\textwidth]{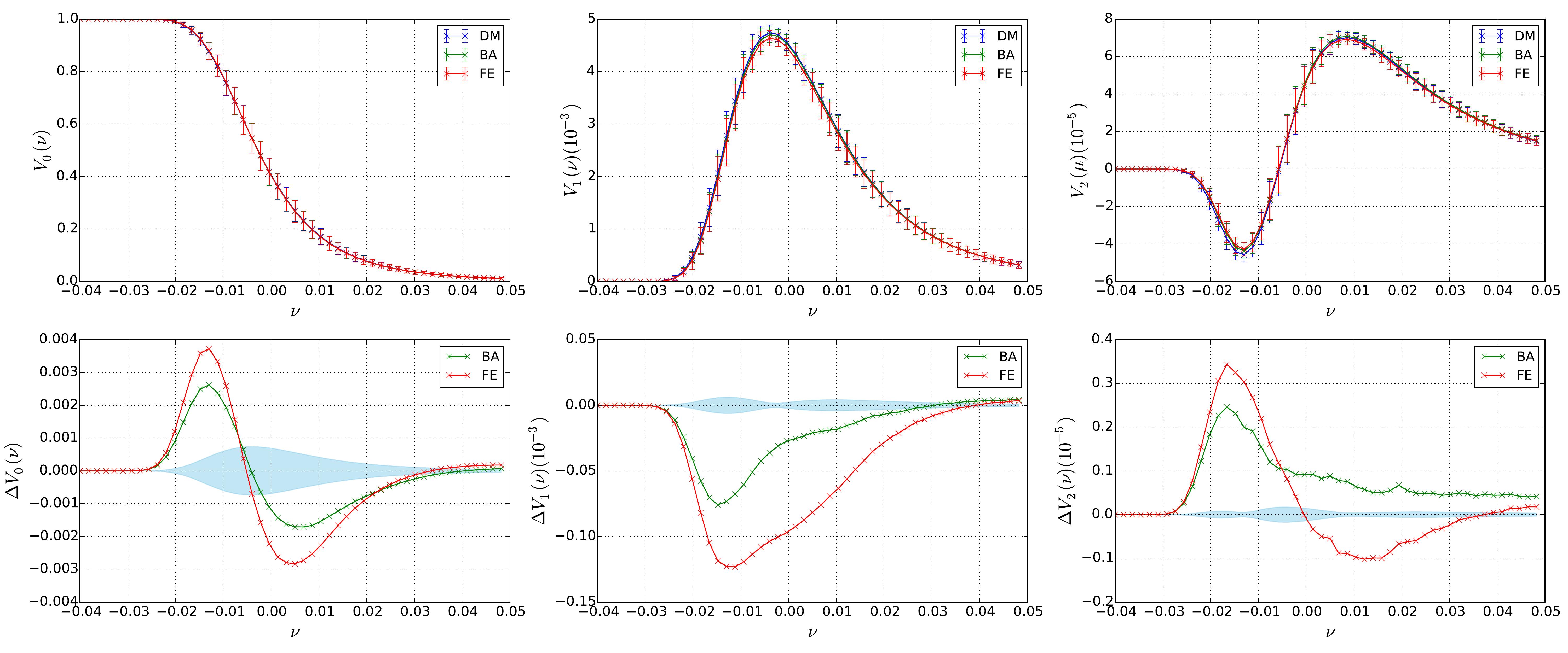}
\centering \includegraphics[clip, width=\textwidth]{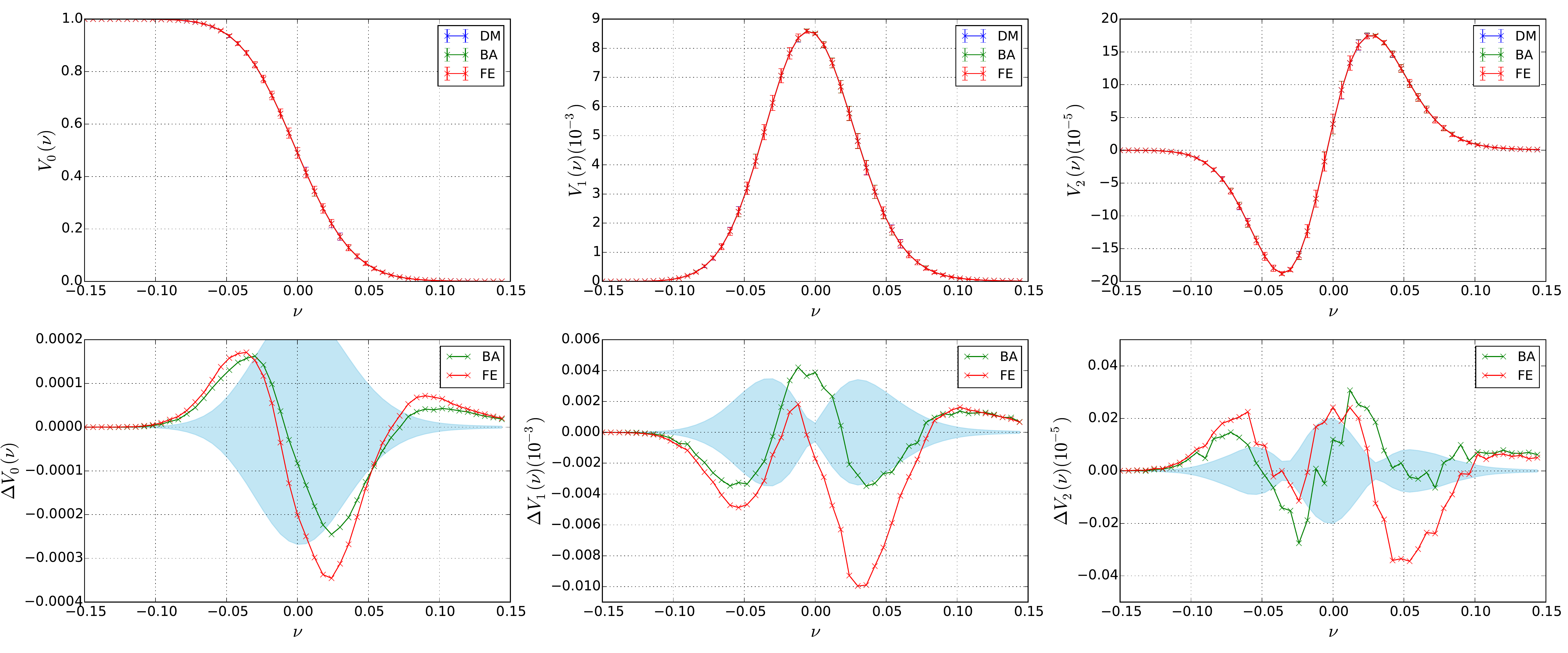}
\caption{The three MFs averaged over one hundred convergence maps. 
Panels in the first and the second rows 
show the results without noise
whereas the third and the fourth rows are for those with noise.
The error bars show the standard deviation of one hundred maps. 
Note that the range of threshold $\nu$ is different 
between the noise-free and noisy cases.
The blue filled region corresponds to
the expected statistical error of DM model from one hundred 1400 deg$^2$ maps.}
\label{fig:minkowski}
\end{figure*}
Figure~\ref{fig:minkowski} show
the MFs computed from our maps of DM, BA and FE models.
Again we use the same color for each model as in the previous figures.
The MFs $V_0$, $V_1$ and $V_2$ are plotted from left to right.
The upper (lower) two rows represent results from noise-free (noisy) maps.
Panels in the second and fourth row 
show the difference of BA and FE models from the DM model.
The error bars represent standard deviations from one hundred maps.
We employ 5 bins per unit $\nu'=(\nu-\langle \mathcal{K} \rangle)/\sigma_0$.
It is useful to calculate the expectation values for a 
Gaussian random field (Eqs.~(\ref{eq:v0_gauss})-(\ref{eq:v2_gauss}))
in order to understand the overall feature of MFs.
In Eqs.~(\ref{eq:v0_gauss})-(\ref{eq:v2_gauss}),
the spectral moments $\sigma_{0}$ and $\sigma_{1}$ can be 
calculated from the convergence power spectrum as
\beqa
\sigma_{p}^2 = \int\frac{{\rm d}^2\ell}{(2\pi)^2}
\ell^{2p}
P_{\kappa}(\ell)
|\tilde{W}_{G}(\ell; \theta_{G})|^2,
\eeqa
where $\tilde{W}_{G}$ represents
the Fourier transform of Eq.~(\ref{eq:gauss_sm}), which is given by
$\exp\left(-\ell^2\theta_{G}^2/4\right)$.
Note that $\tilde{W}_{G}$ decreases exponentially 
at $\ell \simgt 20000 \times (1\, {\rm arcmin}/\theta_{G})$.
>From the result shown in Figure~\ref{fig:spec},
we expect that $\sigma_0$ of BA and FE models is smaller than
that of DM model but $\sigma_1$ would be nearly the same.
We confirm this notion by direct measurement of
$\sigma_0$ and $\sigma_{1}$ from our 100 maps.
It is important to test whether the differences between the models 
are attributed to $\sigma_0$.
In the above analysis, we measured MFs in terms of normalized threshold $\nu'$
because, if the difference is largely owing to variation of $\sigma_0$,
the normalization would absorb all or most of the differences.
The difference in the MFs becomes indeed slightly smaller with the normalization,
but does not vanish completely.
Therefore, the baryonic processes apparently affect the MFs.
It is necessary to develop further analysis beyond Gaussian description
to explain the difference of MFs between baryonic and fiducial models.

\subsection{Fisher forecast and bias of cosmological parameters}
\begin{table}
\centering
\caption{The marginalized errors and biases\label{tb:bias}}
\begin{tabular}{lccc}
\hline
\hline
Data statistics & $\delta \Omega_{\textrm{m}}$ & $\delta w$ & $\delta 10^9 A_s$ \\ \hline
\multicolumn{4}{c}{Fisher forecast marginalized $1\sigma$ error} \\ \hline
PS & $0.0275$ & $0.276$ & $1.01$ \\
MP & $0.00853$ & $0.110$ & $0.0557$ \\
HP & $0.0129$ & $0.118$ & $0.0971$ \\
$V_0$ & $0.00678$ & $0.0711$ & $0.0831$ \\
$V_1$ & $0.00767$ & $0.0911$ & $0.0931$ \\
$V_2$ & $0.00912$ & $0.115$ & $0.0718$ \\ 
$V_0+V_1+V_2$ & $0.00439$ & $0.0513$ & $0.0521$ \\
PS+MP+HP & $0.00478$ & $0.0520$ & $0.0371$ \\
PS+MP+$V_0$ & $0.00372$ & $0.0391$ & $0.0412$ \\
PS+HP+$V_0$ & $0.00547$ & $0.0456$ & $0.0662$ \\\hline
\multicolumn{4}{c}{Parameter bias of BA} \\ \hline
PS & $0.00104$ & $0.000285$ & $-0.0342$ \\
MP & $-0.00280$ & $-0.0394$ & $-0.0187$ \\
HP & $-0.00745$ & $-0.0648$ & $0.0278$ \\
$V_0$ & $-0.00385$ & $-0.0471$ & $0.0435$ \\
$V_1$ & $-0.00349$ & $0.0147$ & $0.101$ \\
$V_2$ & $-0.00518$ & $-0.0479$ & $0.0169$ \\ 
$V_0+V_1+V_2$ & $-0.00271$ & $0.00352$ & $0.0477$ \\
PS+MP+HP & $-0.00115$ & $-0.0145$ & $-0.0134$ \\
PS+MP+$V_0$ & $0.000493$ & $0.00803$ & $0.00428$ \\
PS+HP+$V_0$ & $0.00148$ & $0.0261$ & $0.00488$ \\\hline
\multicolumn{4}{c}{Parameter bias of FE} \\ \hline
PS & $-0.00171$ & $-0.0172$ & $0.0139$ \\
MP & $-0.00591$ & $-0.0616$ & $0.00536$ \\
HP & $-0.00668$ & $-0.0610$ & $0.0347$ \\
$V_0$ & $-0.000951$ & $-0.00112$ & $0.0543$ \\
$V_1$ & $-0.00559$ & $0.00157$ & $0.120$ \\
$V_2$ & $-0.00182$ & $0.0135$ & $0.0342$ \\ 
$V_0+V_1+V_2$ & $-0.000412$ & $0.0347$ & $0.0481$ \\
PS+MP+HP & $-0.00594$ & $-0.0587$ & $0.0156$ \\
PS+MP+$V_0$ & $-0.00288$ & $-0.00899$ & $0.0544$ \\
PS+HP+$V_0$ & $0.000589$ & $0.0136$ & $0.0241$ \\\hline
\end{tabular}
\tablecomments{The marginalized errors of three parameters based on Fisher forecast of DM model and
the biases of BA and FE models compared with the DM model.
Note that these values are not normalized by the fiducial errors.}
\end{table}

\begin{figure*}
\centering \includegraphics[clip, width=0.3\textwidth]{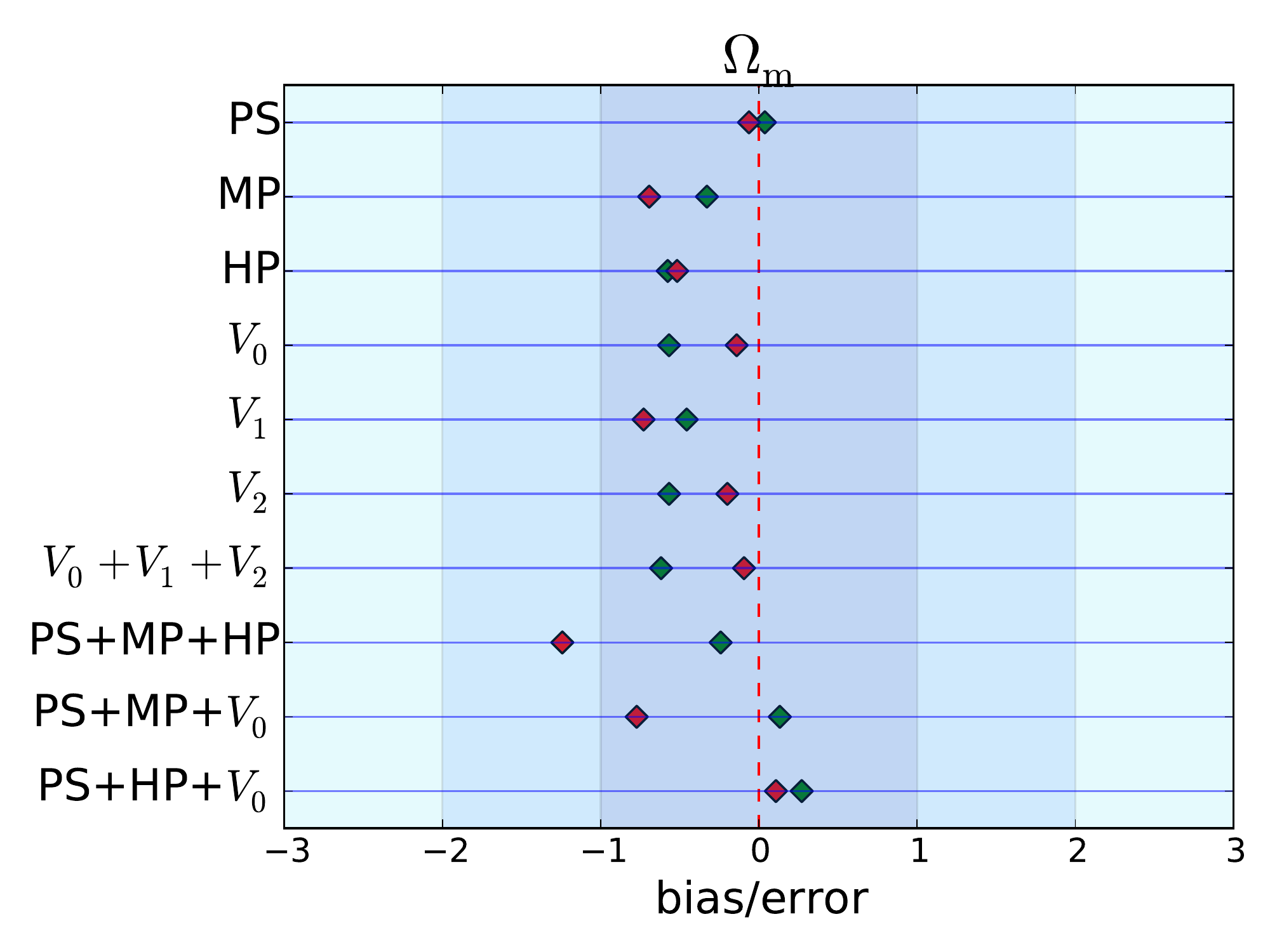}
\centering \includegraphics[clip, width=0.3\textwidth]{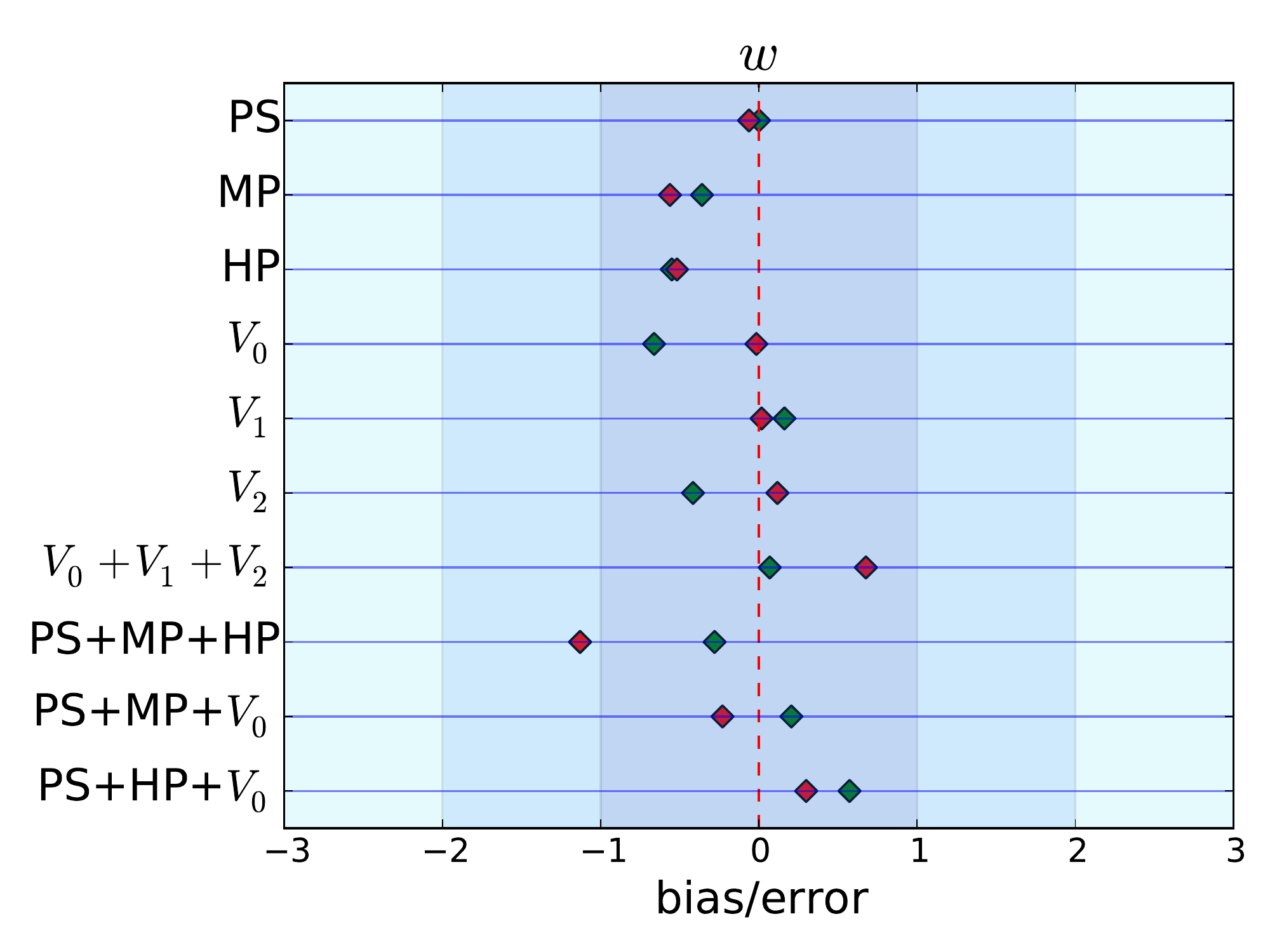}
\centering \includegraphics[clip, width=0.3\textwidth]{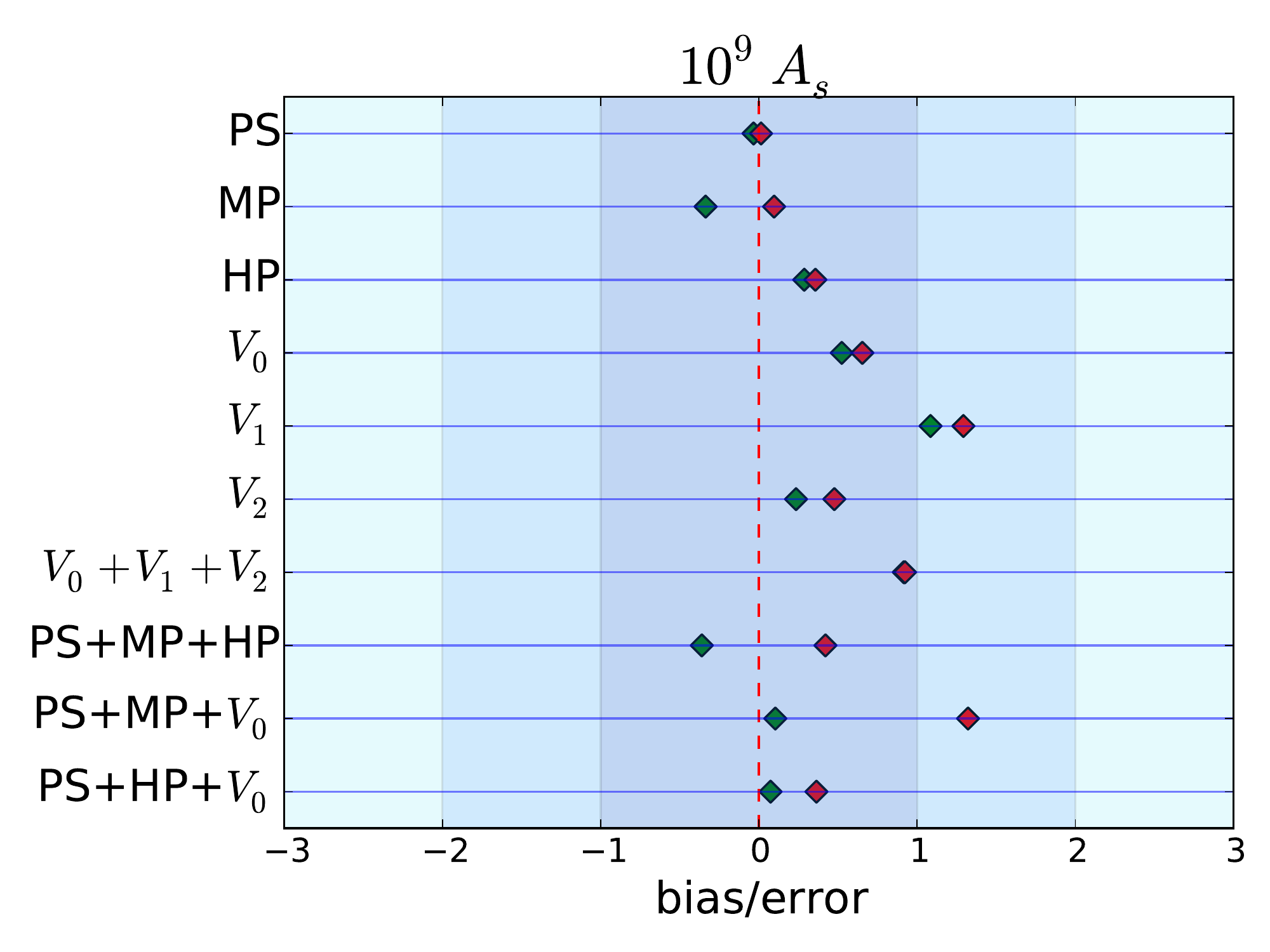}
\caption{The parameter biases divided by marginalized errors. 
Green and red points correspond to BA and FE models respectively. 
The biases caused when using only one observable are mostly within $1\sigma$.
However, in combined analyses, e.g. PS+MP+HP, 
the resulting bias can exceed the marginalized $1\sigma$ error.}
\label{fig:biasratio}
\end{figure*}

\begin{figure}
\centering \includegraphics[clip, width=0.4\textwidth]{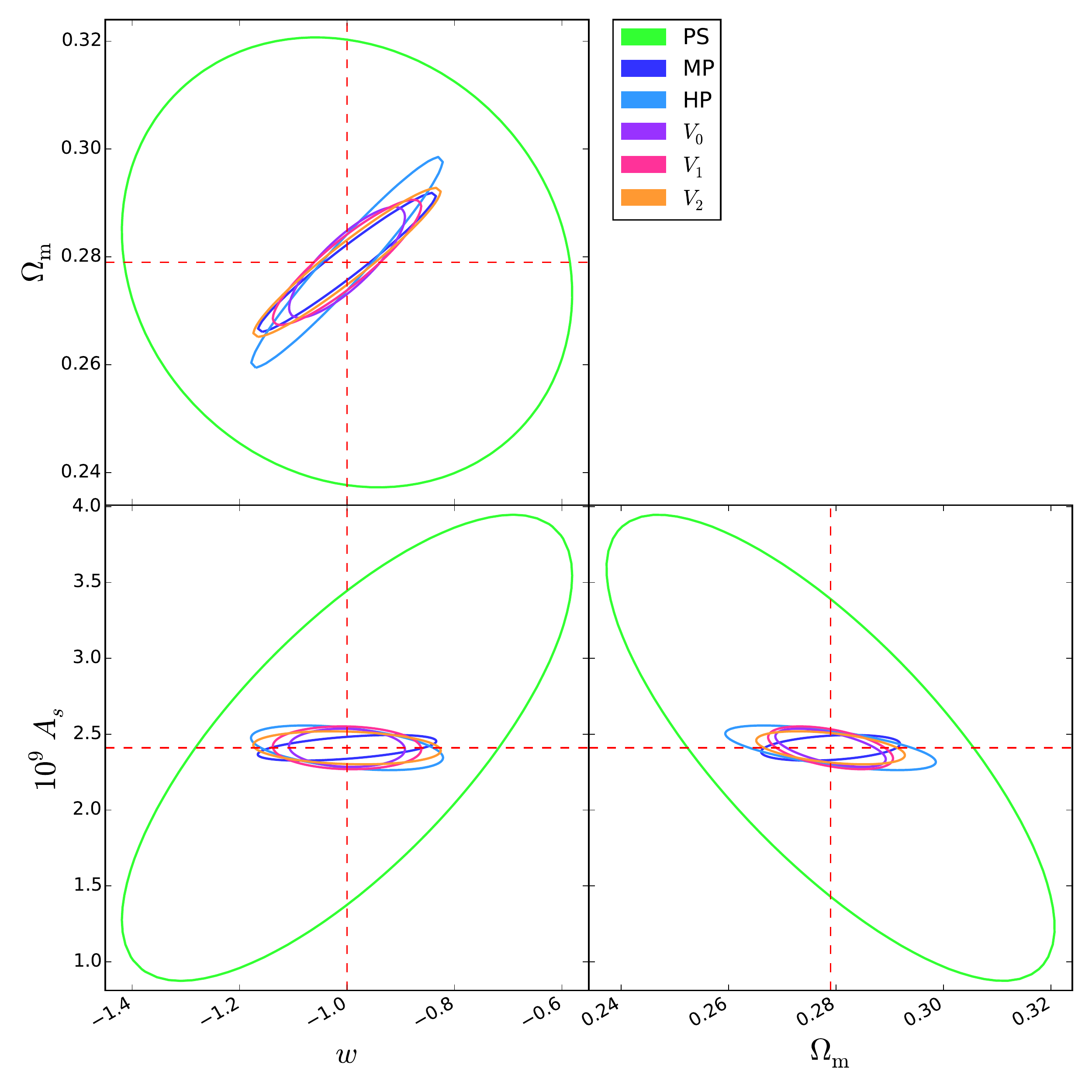}
\caption{We compare the 1$\sigma$ confidence regions calculated from Fisher forecast 
using six observables separately.
The confidence region from PS (green line) is clearly the largest.
The other non-local statistics yield substantially smaller confidence regions.}
\label{fig:error_contour}
\end{figure}

We show the results of our parameter estimation in Figures ~\ref{fig:error_1} 
to \ref{fig:error_3}.
In each figure, we show two-dimensional error contours marginalized over the other parameter. 
The blue, green and red dots indicate
the best-fit values
for DM, BA and FE models, respectively.
Red dashed lines represent the fiducial values.
Note that in these three figures, the plotted parameter range is adjusted
according to the size of the forecast circle for the observable.
In order to make comparison between the figures easy, 
we show the $1\sigma$ confidence regions of six observables
in the same scale in Figure~\ref{fig:error_contour}.
The biases and marginalized errors of parameters
are summarized in Table~\ref{tb:bias} and Figure~\ref{fig:biasratio}.
In Figure~\ref{fig:biasratio}, 
green and red points show 
the biases divided by the marginalized errors for BA and FE models. 
To check the accuracy of our code, 
we also present results from the fiducial simulation.
We plot results obtained from $\cal{K}$ maps with shape noise.

In the following, we summarize the baryonic effects on parameter estimation with
different convergence statistics.

\subsubsection*{Power spectrum}

\begin{figure*}
\centering 
\includegraphics[clip, width=0.8\textwidth]{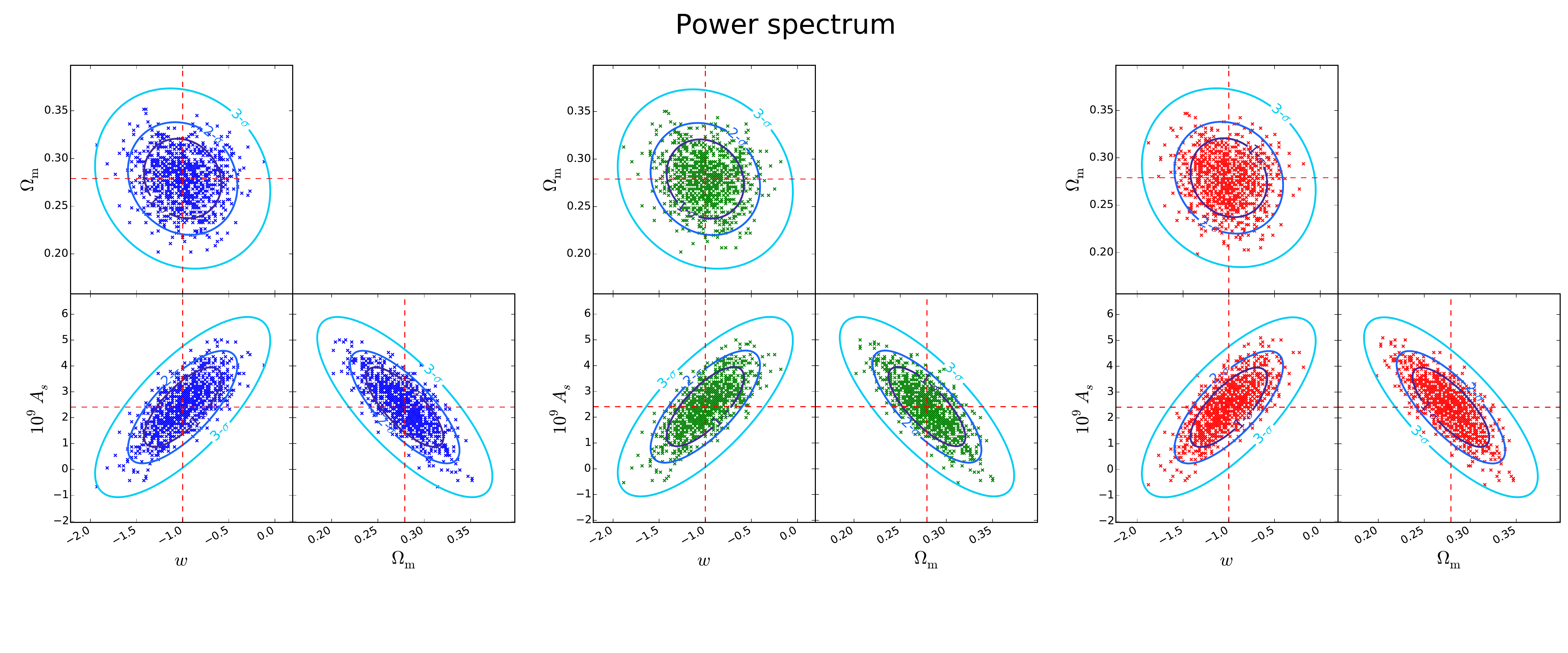}
\centering
\includegraphics[clip, width=0.8\textwidth]{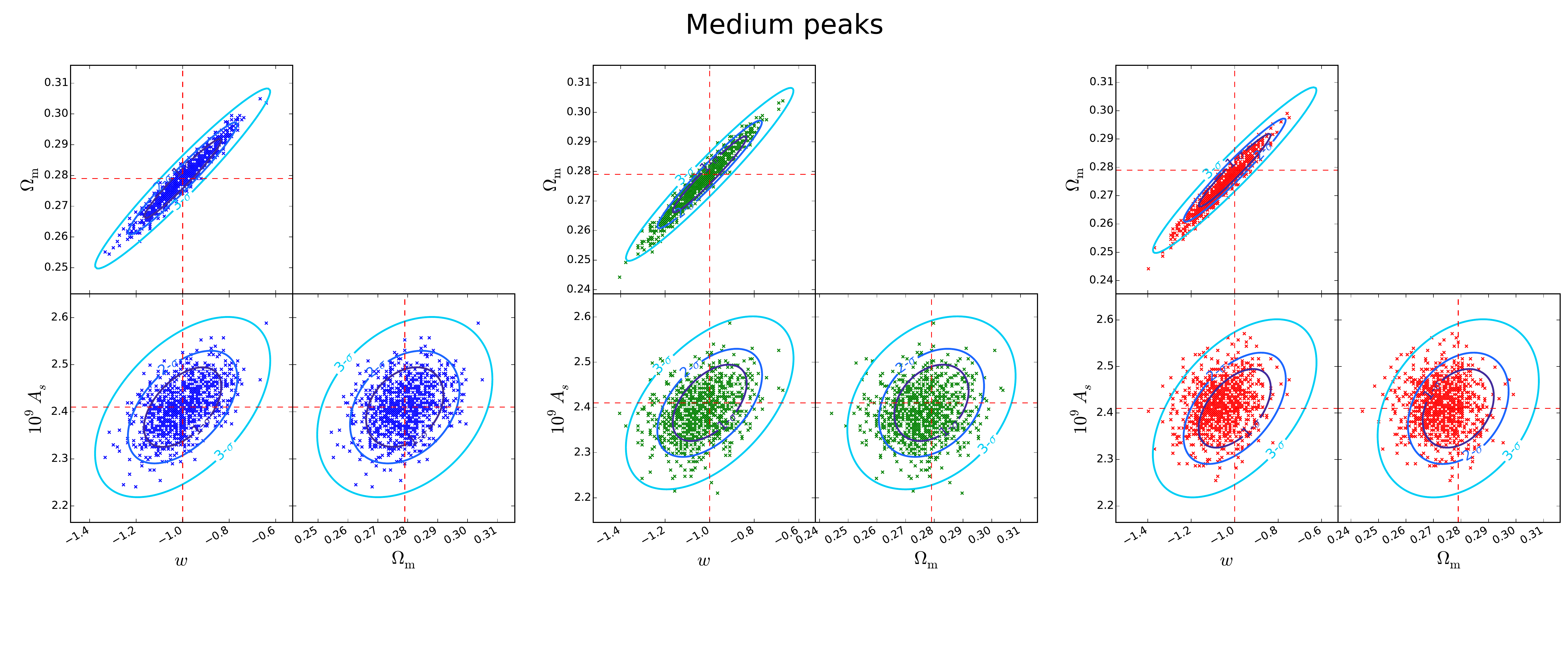}
\centering 
\includegraphics[clip, width=0.8\textwidth]{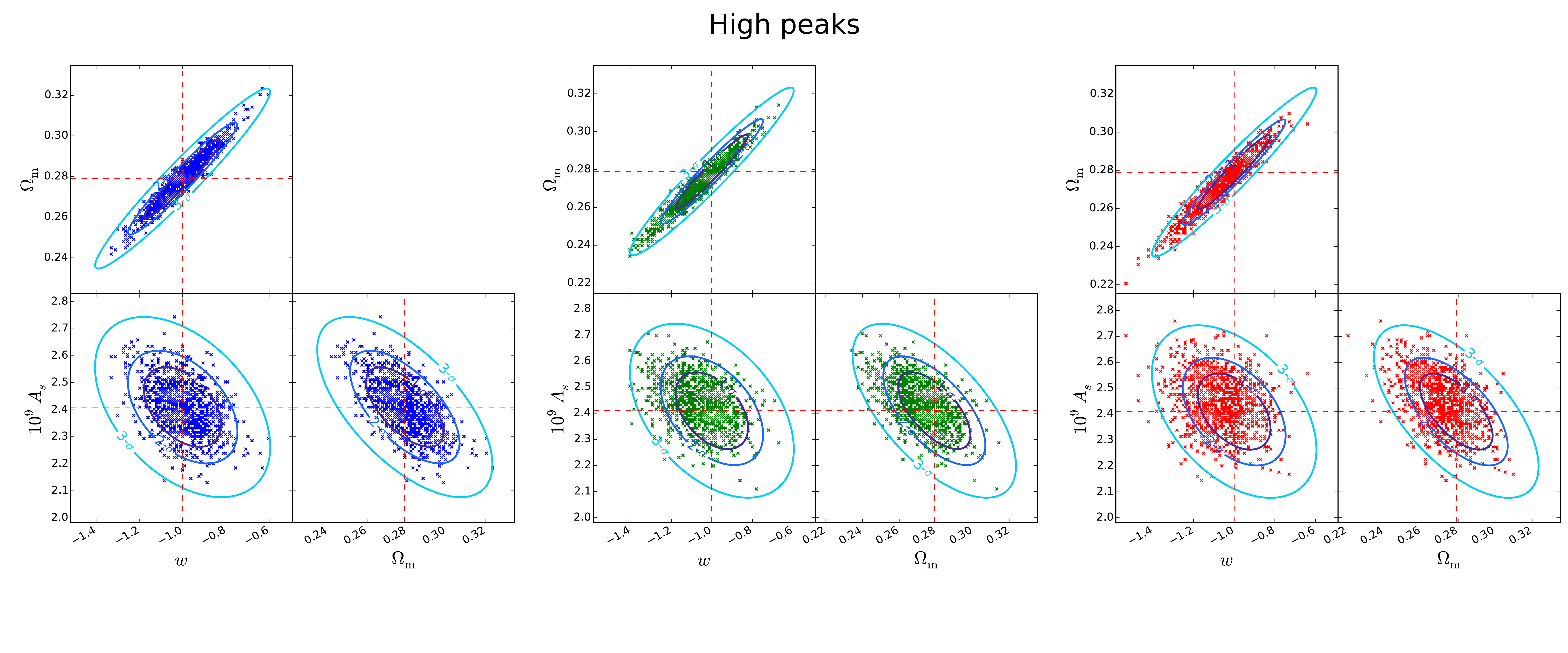}
\caption{Error contours and best-fit points using PS and peak counts. 
Blue, green and red dots indicate the best fit values for each realization of
our DM, BA and FE 
model suites.
Light blue, blue and deep blue contour show  
$1\sigma$, $2\sigma$ and $3\sigma$ confidence boundary
obtained in the Fisher analysis of DM model (described in Sec. \ref{fisher}), respectively.
The discrepancy between the distribution of green or red points
and the blue confidence regions is attributed to ``baryonic effects''.
Red dashed lines represent fiducial parameter values.}
\label{fig:error_1}
\end{figure*}
The top three panels in Figure~\ref{fig:error_1} show the results of
parameter estimation for three (DM, BA, FE) models using only PS.
We find only small bias caused by the baryonic physics. 
That is because, with realistic ground-based surveys 
PS can be measured accurately at low $\ell$ about $2000$ at most, 
which is the maximum multipole used in our estimation. 
The difference between the fiducial and baryonic models appears at the small scale.
The PS amplitude and the shape at $\ell \simlt 2000$ are not significantly 
affected by the baryonic physics.
Previous studies \citep{Yang2013,Mohammed2014} also examined
differences due to the choice of the maximum multipole. 
In our result, the difference at large scale is quite small whereas
noise dominates at smaller angular scales, where PS can not be
measured accurately.
For this reason, we fix the maximum multipole at $\ell =2000$.
Then the baryon effects on PS is 
negligible in cosmological parameter estimation using PS.

\subsubsection*{Peak count}
The medium (bottom) three panels in Figure~\ref{fig:error_1} show
the results using only MPs (HPs).
As we have discussed above, 
medium height peaks are also dominated by intrinsic shape noise
but still have cosmological information.
High peaks have nearly one-to-one relation with massive halos, 
whose masses are less affected by the baryonic processes.
Interestingly, we find that both MPs and HPs induce 
biased parameter estimation of $w$ with a level of $\sim0.56\sigma$ and $0.52\sigma$,
respectively, for FE model.
The bias may be partly originated from the baryonic effect on the shape of massive halos.
Massive halos tend be rounder when baryonic processes are included
\citep[e.g.,][]{Kazantzidis2004}. Then the height of a peak associated
with a single halo is reduced, and so may be the peak counts at
$\cal{K} \sim$ $0.0-0.05$.

\subsubsection*{Minkowski Functionals}
\begin{figure*}
\centering \includegraphics[clip, width=0.8\textwidth]{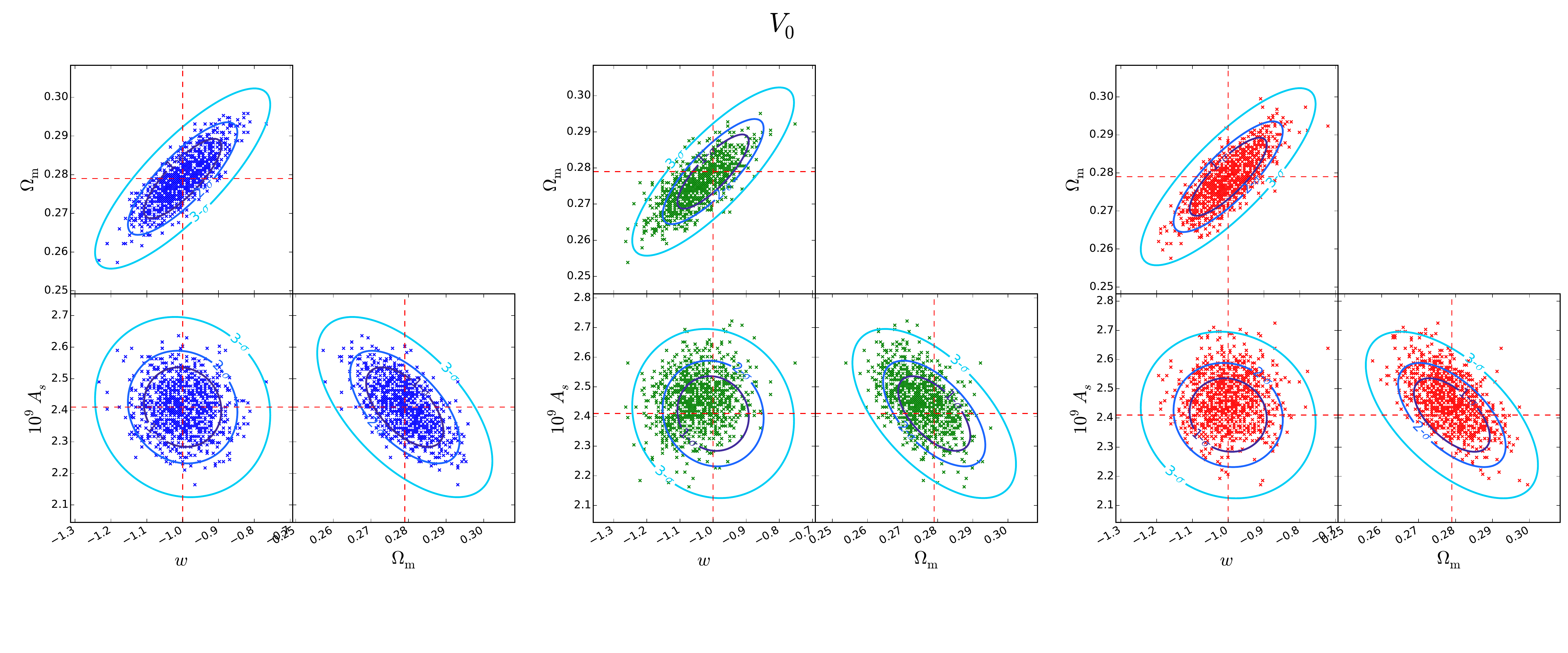}
\centering \includegraphics[clip, width=0.8\textwidth]{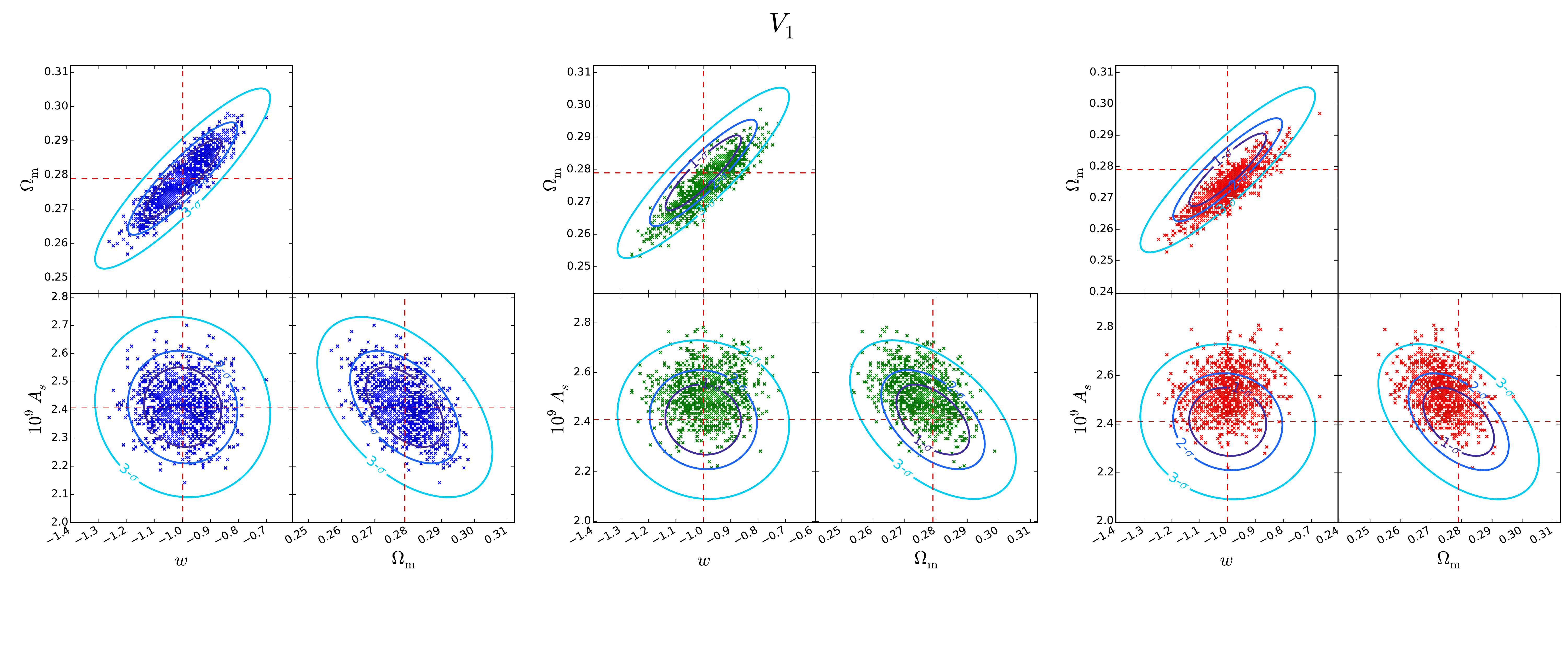}
\centering \includegraphics[clip, width=0.8\textwidth]{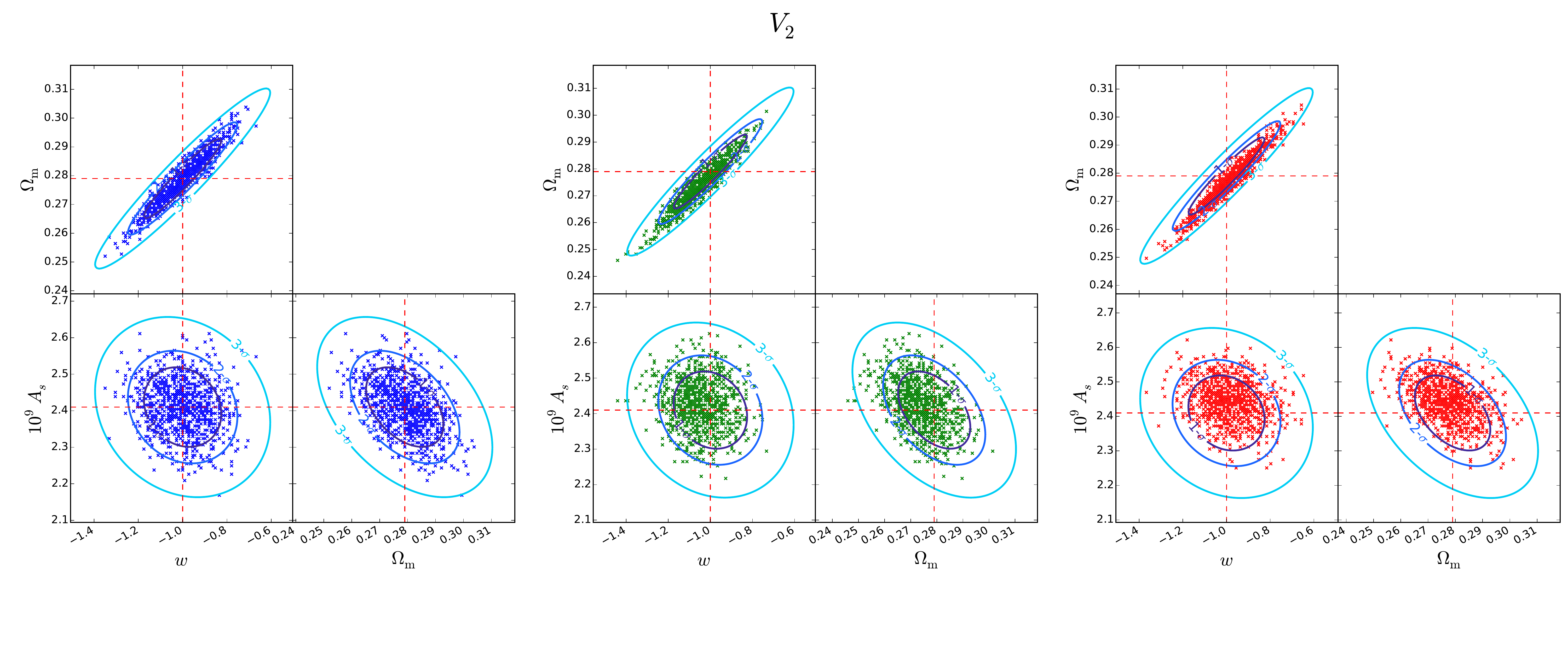}
\caption{Error contours and best-fit points using MFs.}
\label{fig:error_2}
\end{figure*}

Figure~\ref{fig:error_2} shows the results using each of the three MFs,
$V_0, V_1, V_2$. The center of the dots is shifted in the respective
parameter space; significant parameter bias can be caused by the baryonic
effects.
Note that the absolute shift itself is not
very large, but that the bias {\it with respect to the error circle}
is appreciable (see Fig. 8 for the relative size of the error circles). 
When one constructs the theoretical template of MFs without the modeling of baryonic physics,
analyses using $V_0$, $V_1$ and $V_2$ cause the biased estimation of $w$ with a level of 
$\sim0.016\sigma$, $0.017\sigma$, and $0.12\sigma$, respectively, for FE model.
Although it is difficult to explain the origin of this bias completely,
we expect the following two effects can be responsible for the biased parameter estimation:
(i) the change of variance of $\cal{K}$ (i.e. $\sigma_0$) and
(ii) the change in halo shape.
More detailed modeling with an analytic halo approach would be useful to investigate the relation between
the property of dark matter halos and lensing MFs.
This is along the line of our ongoing study using a large set of cosmological simulations.
 
\subsubsection*{Combined analysis}
\begin{figure*}
\centering \includegraphics[clip, width=0.8\textwidth]{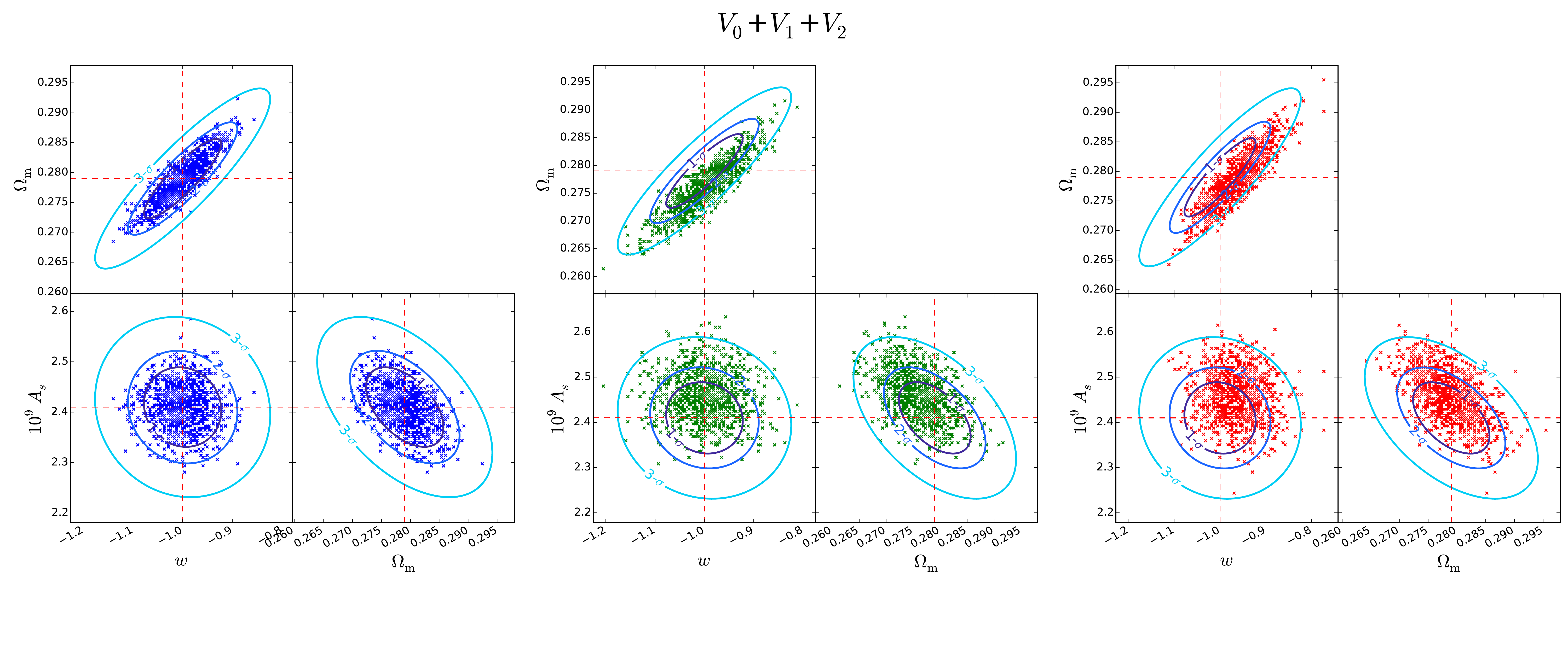}
\centering \includegraphics[clip, width=0.8\textwidth]{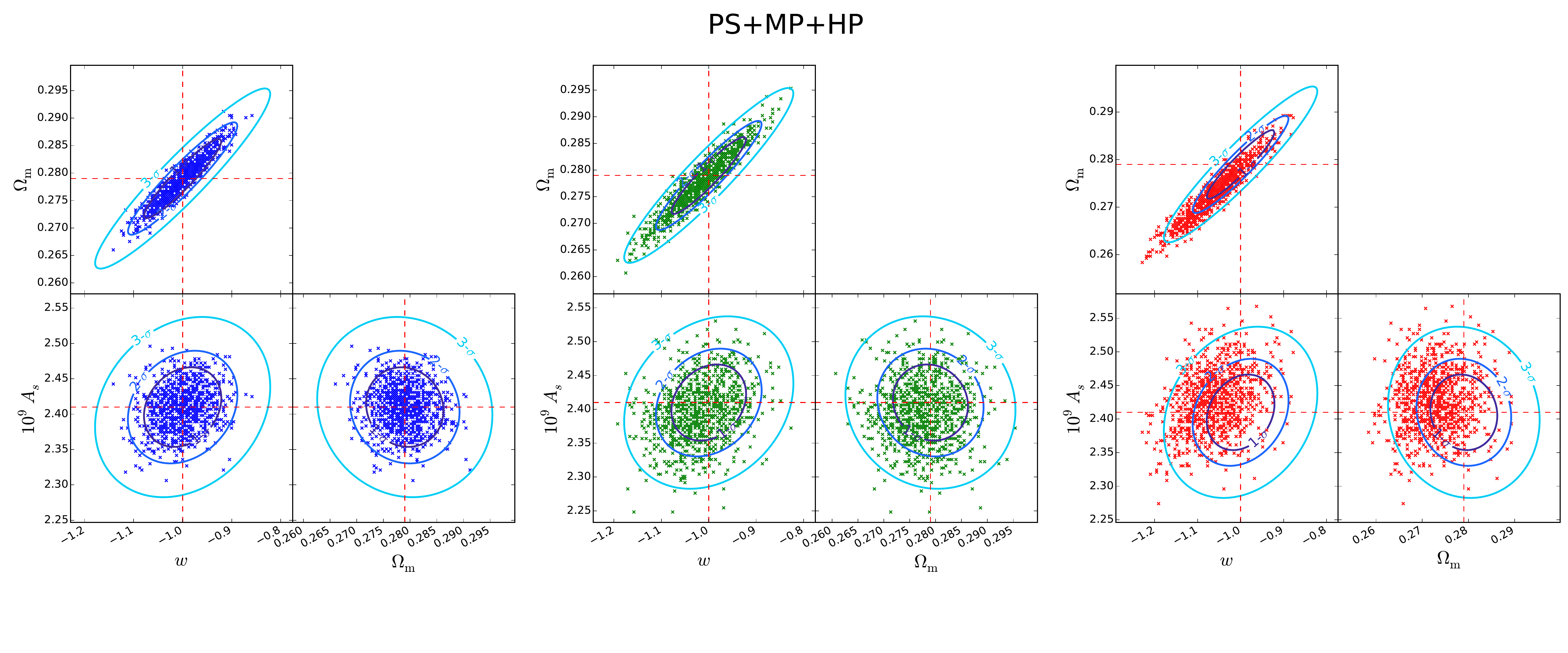}
\centering \includegraphics[clip, width=0.8\textwidth]{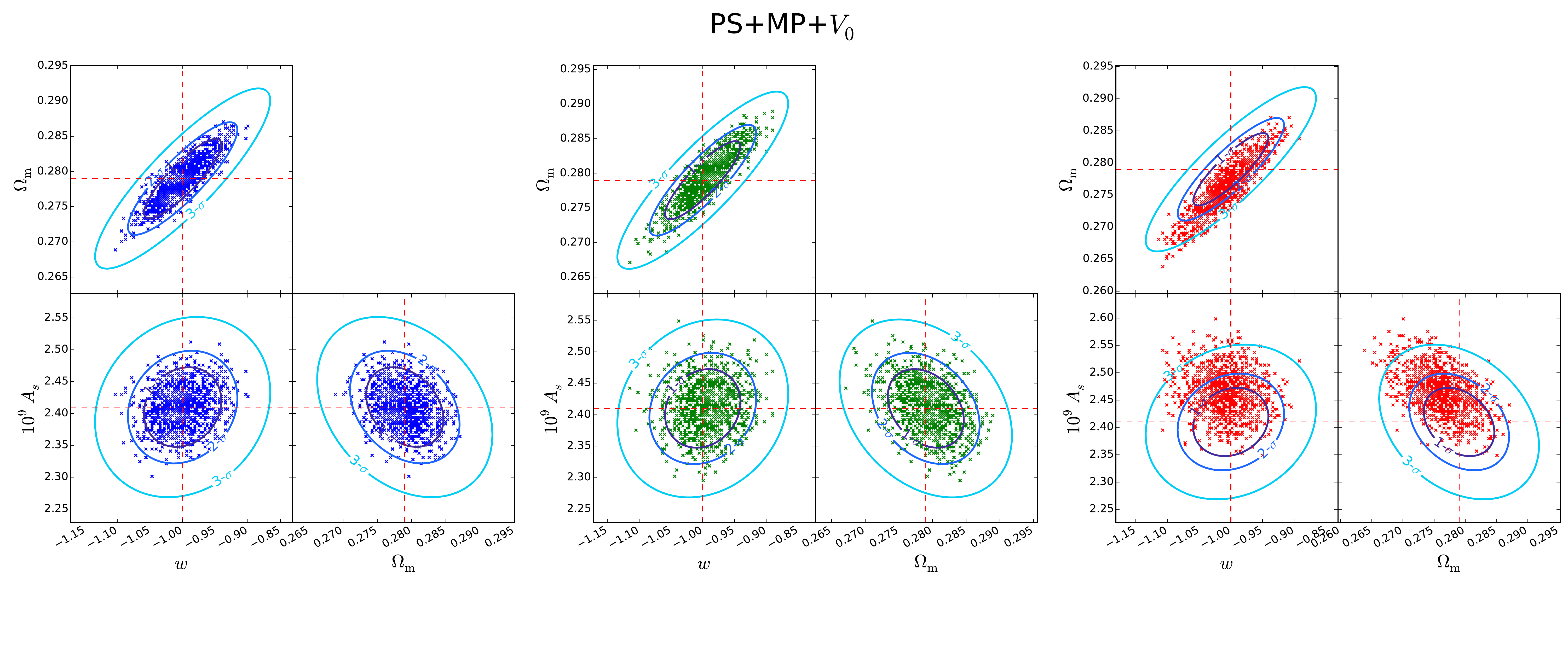}
\centering \includegraphics[clip, width=0.8\textwidth]{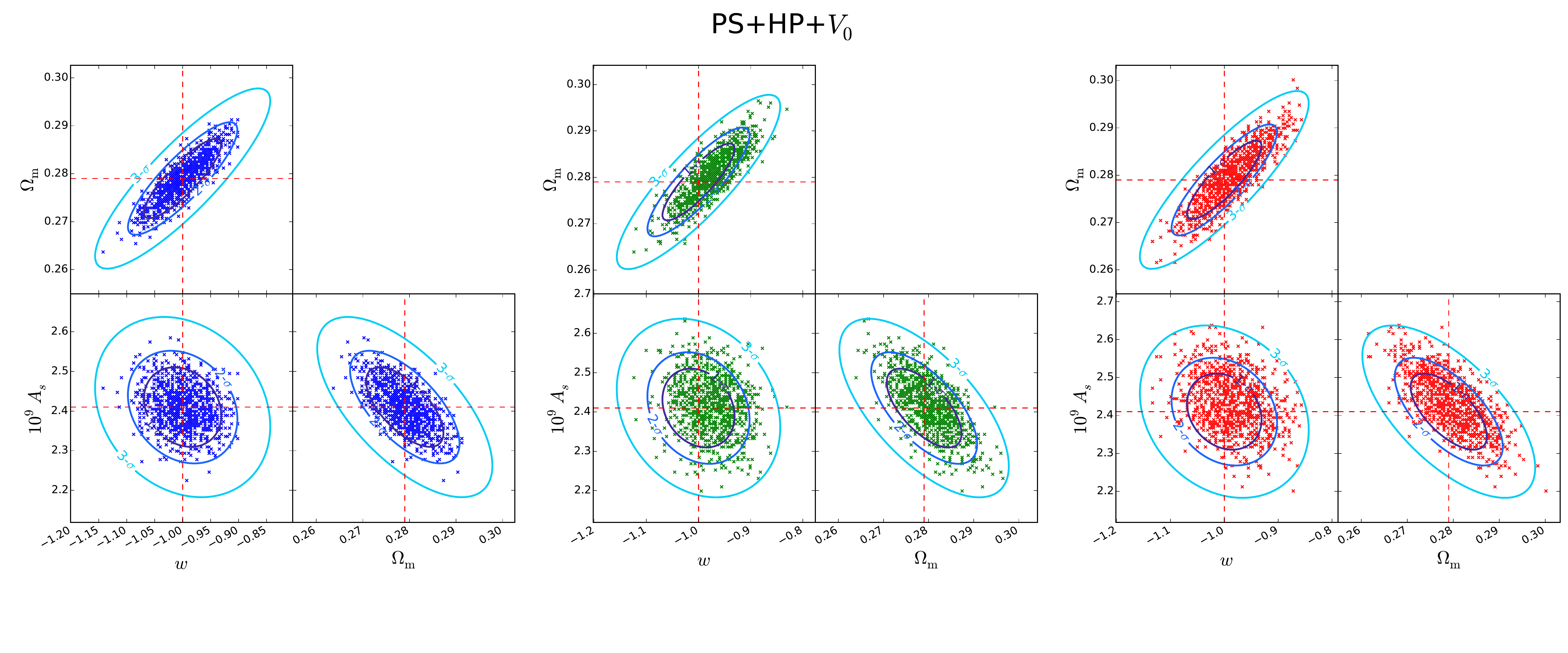}
\caption{Error contours and best-fit points for combined analysis}
\label{fig:error_3}
\end{figure*}
Let us consider a combined analysis with multiple observables 
in order to tighten the error and possibly mitigate the bias due to baryonic effects.

Our combination of the observables are of the following four types:
(i) all the MFs,
(ii) PS and peak counts,
(iii) PS, MPs and $V_0$, and 
(iv) PS, HPs and $V_0$.
The last two are 
examples of \textit{less biased} combinations of 
the statistics.
We propose the two combinations that are expected to cause
small net bias of parameters on the basis of the result of Eq.~(\ref{eq:param_bias}).
The basic idea is to find a combination of observables 
with biases in the opposite directions in parameter space.

Figure~\ref{fig:error_3} shows the results of the above combined analysis.
All of combined analysis presented here 
can tighten forecast errors, 
i.e. they can effectively extract more cosmological information.
Some combinations, e.g. PS and peak counts, 
is largely affected by biases by baryonic effects.
This degrades parameter constraints of combined analysis.
However, if we adopt, e.g. PS+HPs+$V_0$, 
such combination can mitigate the bias by baryonic effects.
When all the statistics cause biases with the same sign (e.g., for $10^9 A_s$),
it is safe to combine those with small biases.
On the other hand, there is still a possibility of 
causing large biases by combined analysis,
if one include statistic(s) whose bias is very large.
In general, cosmological parameters 
are degenerated with each other and thus it is not
trivial to determine the best combination for multiple parameters.
The above case focusing on a single 
may serve as a useful guide for combined analysis when parameter degeneracy 
is not strong.
We further discuss with a simple example with details in multi-dimensional 
parameter space of the formalism in Appendix.


\section{CONCLUSION AND DISCUSSION}
\label{sec:conclusion}

We have studied baryonic effects on WL statistics
using a suite of cosmological simulations 
that incorporate galaxy formation processes.
Various baryonic processes are implemented in our code,
such as gas cooling, star formation, and stellar feedback.
These processes themselves are important subjects of research and
the model uncertainties with free parameters are still controversial,
but such simulations can be used to quantify the baryonic effect, 
or at least to compare the WL statistics with those calculated from 
dark matter only simulations.
We focus on cosmological parameter estimation using WL statistics.
To this end, we made realistic mock observations by
performing ray-tracing simulations through the non-linear density fields
with the size of survey region over 1000 square degrees.

We have studied three statistics, PS, peak counts and MFs,
calculated directly from mock lensing maps with baryons and compare them 
with the results from our `fiducial' dark matter only simulations.
The PS deviate appreciably at $\ell \simgt 7000$ from the fiducial model 
due to the baryonic processes.
This feature is also seen in previous studies, e.g., \citet{Semboloni2011}.
The shape noise dominates, however, over the baryonic effects at the
small angular scales and
thus the baryonic effects are not critical in practice in the analysis using PS.
Peak counts in convergence maps are also affected because the height 
of a peak is sensitive to the mass and the density profile of the corresponding halo.
Stellar feedback can effectively reduce the mass of small galaxy halos, 
which results in decrease of the number of medium height peaks.
High peaks are less affected by the stellar feedback effects because 
the mass distribution in massive, cluster-size halos are not significantly changed
by the stellar feedback.
\citet{Yang2013} examine the influence on peak counts
of the enhancement of halo concentration parameter,
motivated by radiative cooling.
They show that the number of HPs is increased accordingly,
while the number of MPs is mostly unaffected.
We do not see the feature in our high peak-counts, 
probably because our simulations include
not only radiative cooling but also supernova and stellar feedback 
that can compensate, at least partially, the condensation of baryons.
Finally, MFs are morphological statistics and thus are promising as a probe
of the baryonic effects that change the internal matter distribution of halos. 
The shape noise again substantially affects MFs owing to the effectively increased
$\kappa$-variance. Also, the smoother distribution of the gaseous components
causes the overall feature of the MF smoother, especially in our BA model (see Fig. 6).

We have considered three kinds of statistics as probes of cosmological 
parameters. When we use only one of them,
the parameter bias due to the baryonic effects is not significant.
Most of the biases are found to be within 1$\sigma$ error
for the fiducial 1400 square degree survey. 
It is expected that, with the upcoming Subaru HSC survey,
cosmological parameters can be determined
without being significantly compromised by baryonic processes
as long as a single statistic is used.
However, because the statistical error itself becomes small, 
roughly in proportion to the increase
of square root of the survey area, even a very small bias would become critical
for future surveys with a half or an all sky
coverage.
The overall bias can appear relatively amplified
when a combination of a few or more statistics
are used because the expected error of the parameters 
becomes small and the bias remains the same.
For example, when using both PS and
peak counts, the parameter bias for $w$ is over $1\sigma$. 
Clearly, such bias needs to be well understood before analyzing real observational data.
It is also desirable to find a combination of statistics that yields both high 
precision and small bias.
If we consider only one parameter, a best way would be to combine statistics such that
their respective biases can cancel each other (see Figure 7).
Unfortunately, it is generally non-trivial to estimate parameter bias
in a multi-dimensional space
because the degeneracy between multiple parameters leads to
complicated dependence on parameters.
We suggest to try all possible combinations and study in detail,
as has been done in the present paper 
in the case with only a few statistical measures.
Among the combinations we have tested, we find the combination of PS, 
one of peak counts and $V_0$
gives high precision and yet robust results against the baryonic effects.
Furthermore, all of the quantities can be predicted accurately by analytic models.
\citet{Takahashi2012} studied non-linear PS using HaloFits approach.
\citet{Hamana2004} employs one-to-one relation that NFW-halo corresponds to a peak
but this approach is only valid at high S/N ratio ($\simgt 4$) peaks.
For low S/N ratio peaks, this relation breaks down.
\citet{Das2006} studied the probability distribution function of convergence
using fitting formula motivated by the log-normal distribution.
This probability distribution function is directly related with $V_0$.
In recent studies, \citet{Fedeli2014a,Fedeli2014b,Mohammed2014}
study baryonic effects on PS using a halo model.
It would be interesting to study the link between other two statistics and properties of halos 
by extending the halo model approach. We leave it as a future work. 
The connection between halo properties and WL statistics is also interesting
in that we can possibly observe halo properties through WL statistics.

Our results suggest that the non-local statistics have a stronger
parameter constraining power than PS,
even when baryonic effects are taken into account.
However, one needs to be careful in using the non-local statistics.
Theoretical frameworks to predict PS have been well developed
through both numerical and analytic approach.
In realistic observations, the shear measurement itself is clean and local,
especially when compared with the other non-local statistics.
Overall, PS is the most studied statistic and it
is intuitively easy to understand based on physics; 
$P_\kappa (\ell)$ is a direct measure of the fluctuations at a given scale $\ell$.
Although some numerical studies \citep{Kratochvil2012,Shirasaki2014} suggest that 
non-local statistics would be useful to make tight cosmological constraints,
there are intrinsic difficulties to derive analytical models 
for MFs and peak counts. Hence it is not straightforward  
to interpret the non-local statistics and their dependence on 
cosmological parameters.
In order to rely on the non-local statistics with real data,
full comprehension of systematic uncertainties 
induced by observational effects is necessary. 
Clearly, these issues are worth studying further
in order to explore fully the applicability 
of the non-local statistics.
In this paper, we have clarified the theoretical uncertainty 
of non-local statistics due to baryonic processes.
Our result is a key to the precise application of non-local 
statistics of cosmic shear for cosmological analyses
with the unprecedented large surveys.


\acknowledgments
The authors are grateful to Ikkoh Shimizu for providing codes and for help
with galaxy formation simulations,
and to Zolt\'an Haiman for useful comments on the earlier version of our manuscript.
We appreciate constructive comments of the referee and Joachim Harnois-Deraps.
KO is supported by Advanced Leading Graduate Course for Photon Science.
MS is supported by Research Fellowships of the Japan Society
for the Promotion of Science (JSPS) for Young Scientists.
NY acknowledges financial support from JST CREST 
and from the Japan Society for the Promotion of Science (JSPS) 
Grant-in-Aid for Scientific Research (25287050).
Numerical simulations were carried out on Cray XC30 at Center for Computational Astrophysics,
National Astronomical Observatory of Japan.

\appendix
\section{Combined analysis in one-dimensional parameter space}
\label{sec:appendix}
We present the criterion for the choice of observables for a combined analysis in a simplified case.
The parameter bias of a combined analysis is given by
\beqa
\delta \hat{p}_\alpha = (\hat{F})^{-1}_{\alpha \beta} \sum_{i,j} (N_i-\bar{N}_i)(\hat{C})^{-1}_{ij} \frac{\partial N_j}{\partial p_\beta},
\eeqa
where $\hat{F}$ and $\hat{C}$ is the Fisher matrix and the covariance matrix
in the case of combined analysis with several observables. 
\footnote{In this appendix, we always take sum for greek letters.}
Assuming that each observable is independent, 
one can write $\hat{p}_\alpha$ as
\beqa
\delta \hat{p}_\alpha 
&=& (\hat{F})^{-1}_{\alpha \beta} 
\sum_{i,j,s} (N^s_i-\bar{N}^s_i)(C^s)^{-1}_{ij} \frac{\partial N^s_j}{\partial p_\beta} \\
&=& (\hat{F})^{-1}_{\alpha \beta} \sum_s (F^s)_{\gamma \beta} \delta p _\gamma^s,
\eeqa
where $F^s$ and $C^s$ is the Fisher matrix and the covariance matrix 
for each observable $s$, e.g. $s=$PS. 

\begin{figure*}
\centering \includegraphics[clip, width=0.3\textwidth]{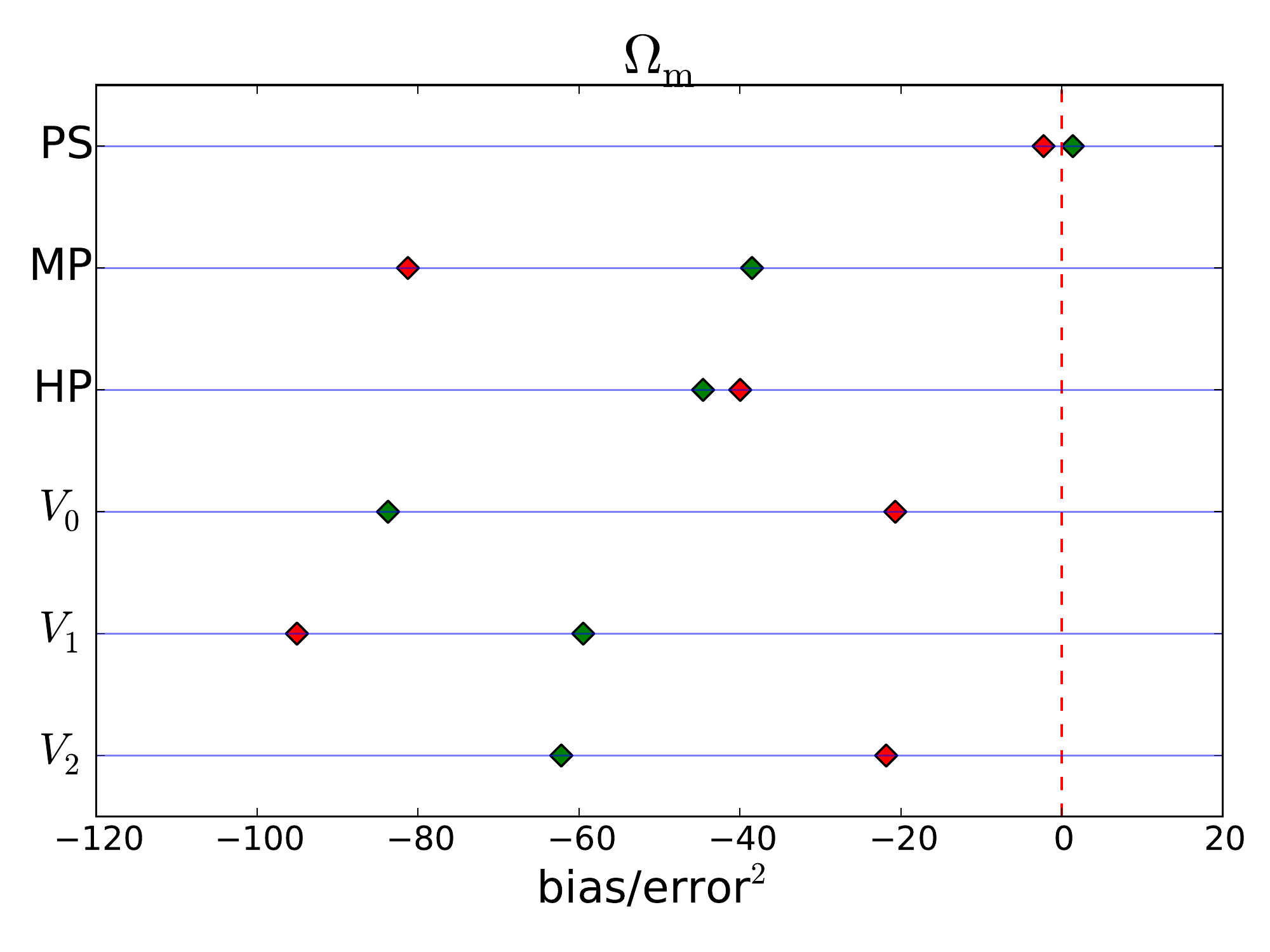}
\centering \includegraphics[clip, width=0.3\textwidth]{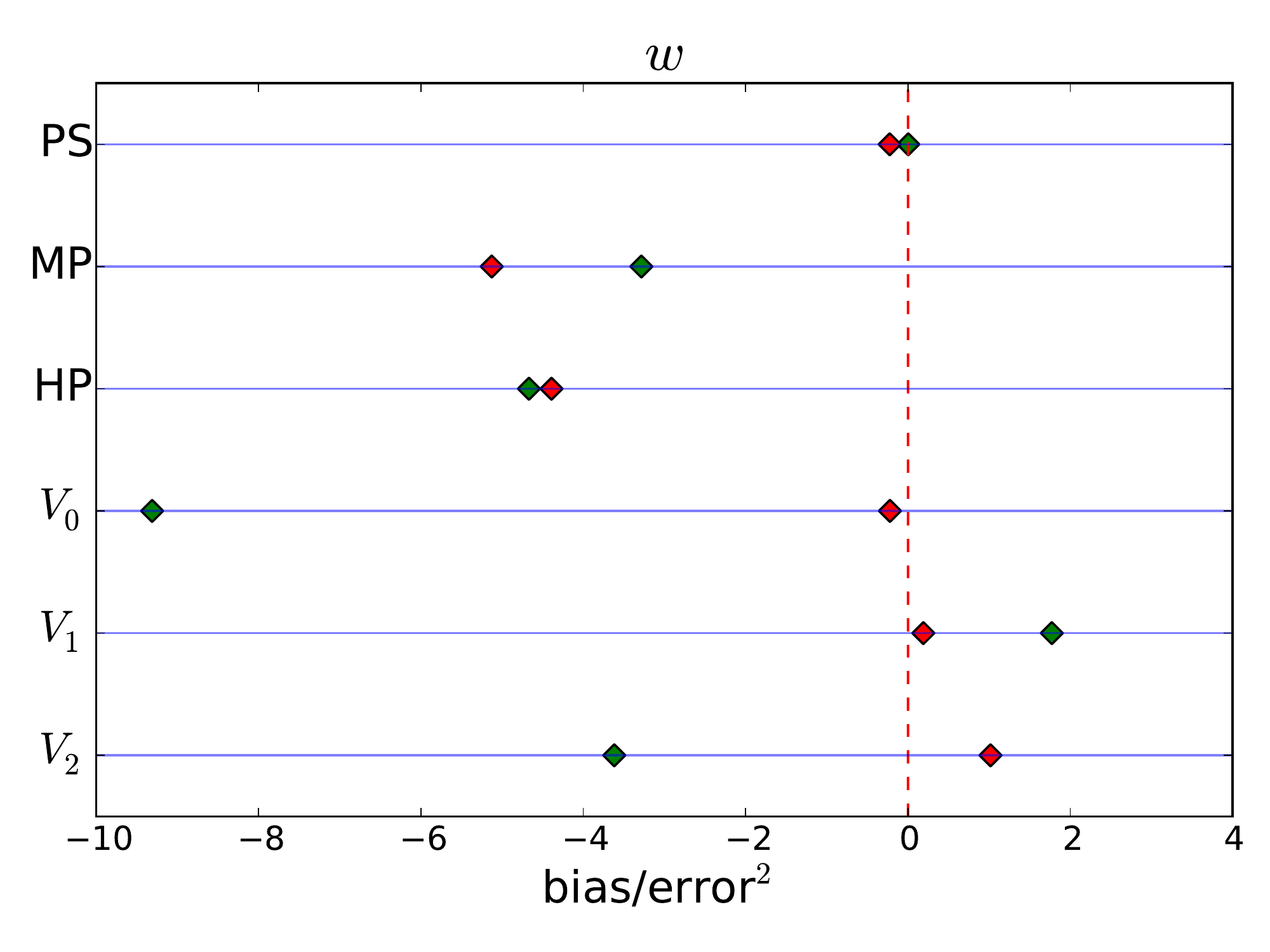}
\centering \includegraphics[clip, width=0.3\textwidth]{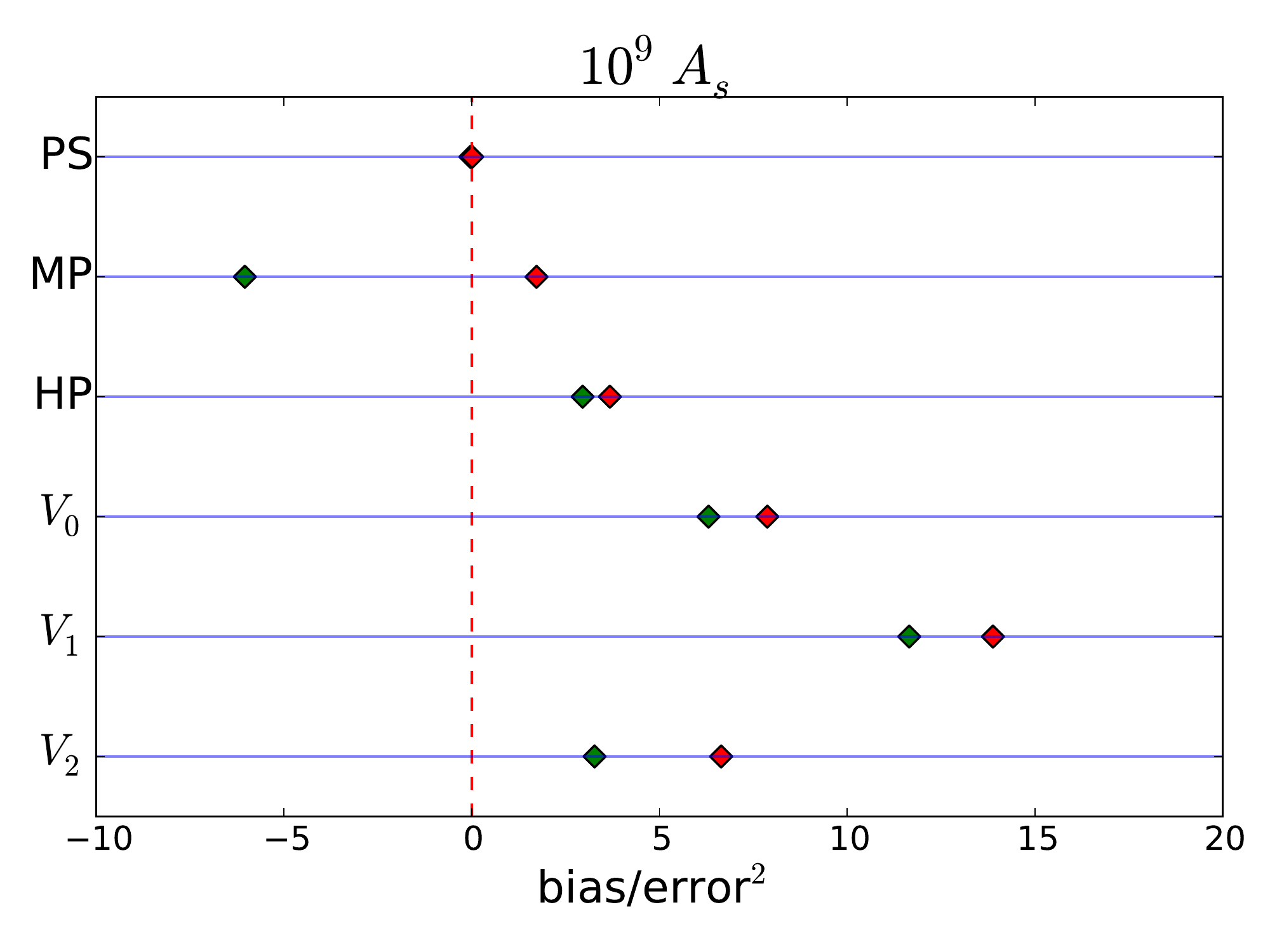}
\caption{The parameter biases divided by the square of marginalized errors. Green and red points correspond to BA and FE models respectively.}
\label{fig:biassquare}
\end{figure*}

For one-dimensional parameter space, we can simplify Fisher matrices,
\beqa
F^s \to \sigma_s^{-2},\ \hat{F}^s \to \sum_s \sigma_s^{-2},
\eeqa
where $\sigma_s$ is the forecast error with the observable $s$. 
Hence, the expected bias with a combined analysis is given by
\beqa
\label{eq:joint_bias}
\delta \hat{p} = \left( \sum_{s'} \sigma_{s'}^{-2} \right)^{-1} \sum_s \frac{\delta p^s}{\sigma_s^2}.
\eeqa
When the sum of $\delta p^s/\sigma_s^2$ vanishes, the net bias $\delta \hat{p}$ also vanishes.
Therefore, a combination that makes the sum small is a good probe with less bias.
We show the bias divided by the square of the error for each observable in Figure~\ref{fig:biassquare}.
For multi-dimensional parameter space, this procedure is not generally
applicable because the degeneracy between 
different parameters gives additional terms to Eq.~(\ref{eq:joint_bias}). 
But when the degeneracy is not strong and the bias is not large, 
this condition gives the effective criterion for a \textit{less biased} combination of observables
in terms of cosmological parameters.

\bibliography{bibtex_library}

\end{document}